\documentclass[journal=jpcafh,manuscript=article,layout=traditional]{achemso}

\pdftrailerid{}
\usepackage[T1]{fontenc} \usepackage{mathtools}
\usepackage{graphicx}
\usepackage{lineno}
\usepackage{amsmath}
\usepackage{amssymb}
\usepackage{physics}
\usepackage{mathrsfs}
\usepackage{bm}
\usepackage{threeparttable}
\usepackage{float}
\usepackage{longtable}
\usepackage{multirow}
\usepackage{booktabs}
\usepackage{hyperref}
\usepackage[section]{placeins}
\usepackage{makecell}
\usepackage{xcolor}

\newcommand{\recCu}{46.91 \pm 1.31}
\newcommand{\recAg}{50.97 \pm 1.93}
\newcommand{\recAu}{36.68 \pm 0.78}

\newcommand{\recCuSOC}{0.39}
\newcommand{\recAgSOC}{0.18}
\newcommand{\recAuSOC}{0.11}

\newcommand{\recCuErrorDeltaT}{2.46}
\newcommand{\recAgErrorDeltaT}{3.66}
\newcommand{\recAuErrorDeltaT}{1.17}
\newcommand{\recCuErrorNoFit}{1.29}
\newcommand{\recAgErrorNoFit}{1.92}
\newcommand{\recAuErrorNoFit}{0.78}

\SectionNumbersOn

\makeatletter
\newcommand{\warninput}[1]{\filename@parse{#1}\InputIfFileExists{#1}{}{\message{LaTeX Warning: File `\filename@base.\ifx\filename@ext\relax tex\else\filename@ext\fi' not found on input line \the\inputlineno}}}
\makeatother

\title{Relativistic and electron-correlation effects in static dipole polarizabilities for group 11 elements}

\author{YingXing Cheng}
\affiliation[University of Stuttgart]{
Institute of Applied Analysis and Numerical Simulation, University of Stuttgart, Pfaffenwaldring 57, 70569, Stuttgart, Germany
}
\email{yingxing.cheng@mathematik.uni-stuttgart.de}

\begin{document}

    \begin{abstract}
        The static dipole polarizabilities of group 11 elements (Cu, Ag, and Au) are computed using the relativistic coupled-cluster method with single, double, and perturbative triple excitations.
Three types of relativistic effects on dipole polarizabilities are investigated: scalar-relativistic, spin-orbit coupling (SOC), and fully relativistic Dirac-Coulomb contributions.
The final recommended values, including uncertainties, are $46.91 \pm 1.31$ a.u. for Cu, $50.97 \pm 1.93$ a.u. for Ag, and $36.68 \pm 0.78$ a.u. for Au.
Our results show close agreement with the values recommended in the 2018 Table of static dipole polarizabilities for neutral elements [\textit{Mol. Phys.} \textbf{2019}, \textit{117}, 1200], with reduced uncertainties for Ag and Au.
The analysis indicates that scalar-relativistic effects are the dominant relativistic contribution for these elements, while SOC effects are negligible.
The influence of electron correlation across all relativistic regimes is also evaluated, demonstrating its significant role in the accurate calculation of dipole polarizabilities.
     \end{abstract}

    \newpage

    \section{Introduction}
    \label{sec:introduction}
    The electric dipole polarizability quantifies the deformation of a system's electron density in response to an external electric field, giving rise to induced dipoles.
Accurately determining static dipole polarizabilities is essential for understanding fundamental interactions in atomic and molecular physics, such as atomic scattering cross sections, refractive indices, dielectric constants, interatomic interactions, and the development of polarizable force fields in molecular simulations \cite{Schwerdtfeger2019}.
Reliable dipole polarizability values also provide crucial benchmarks for methods like density-functional theory \cite{Bast2008}, aid in the development of new basis sets in computational chemistry, such as the Dyall-family basis sets \cite{Dyall2011, Dyall2016, Dyall2023}, and help validate emerging basis sets derived from the Douglas-Kroll-Hess formalism \cite{Ferreira2020} and zeroth-order regular approximation basis sets \cite{CanalNeto2021, Neto2021, Centoducatte2022, Neto2023, Gomes2024, Sampaio2024}.

Moreover, precise knowledge of atomic dipole polarizabilities is critical for improving the accuracy of high-precision atomic clocks based on optical transitions.
The cesium ($Z=55$) atomic clock, which measures the frequency of microwave radiation associated with the transition between two hyperfine levels of the Cs ground state, defines the second, the base unit of time in the International System of Units (SI) \cite{Newell2019}.
This precise frequency standard is essential not only for maintaining global time standards but also for navigation systems, such as the Global Positioning System (GPS), where accurate timekeeping is crucial for determining precise positions \cite{Major2007, Ludlow2015}.

In recent years, atomic clocks have also been used to explore new physics, such as the potential variation of fundamental constants over time \cite{Fischer2004, Blatt2008, Rosenband2008, Dzuba2021}.
For instance, Dzuba \textit{et al.} investigated copper ($Z=29$), silver ($Z=47$), and gold ($Z=79$) clocks, based on transitions from the ground state to the metastable $^2D_{5/2}$ state, to evaluate their potential as optical lattice clocks.
Their study showed that Au clocks, in particular, are highly sensitive to variations in the fine-structure constant, scalar dark matter, and violations of local Lorentz invariance (LLI), while Cu and Ag are also promising candidates for LLI tests \cite{Dzuba2021}.

A significant challenge in advancing higher-precision atomic clocks is mitigating the effects of the black-body radiation (BBR) shift, where the leading-order shifts are proportional to the differential polarizability between the two clock states \cite{Ludlow2015}.
Thus, improving the precision of static polarizability measurements directly enhances the accuracy of atomic clocks \cite{Safronova2012a}.

Furthermore, accurate static polarizability values are essential for evaluating excited-state polarizabilities used in high-order many-body methods like CI+MBPT \cite{Kozlov2001,Porsev2006,Porsev2006a,Dzuba2009,Mitroy2010a} and all-order many-body method \cite{Blundell1989,Derevianko1999,Derevianko2010,Safronova2007,Johnson2008,Badhan2022}, which optimize correlation treatments.
These values also play a vital role in mitigating the BBR shift in atomic clocks \cite{Safronova2012a}, enhancing the accuracy of timekeeping systems and facilitating tests of fundamental physics.

While Cheng recently computed dipole polarizabilities for main-group elements, excluding hydrogen \cite{Cheng2024a}, using fully relativistic Dirac-Coulomb coupled-cluster (CC) and configuration interaction (CI) methods with extensive Dyall \cite{Dyall2002, Dyall2004, Dyall2006, Dyall2007, Dyall2009, Dyall2010, Dyall2011, Dyall2016} and ANO-RCC basis sets \cite{Roos2004, Roos2005}, no comprehensive study to date has applied the CC method within a fully relativistic Dirac-Coulomb framework to group 11 elements, including Cu, Ag, and Au.
This work aims to fill that gap.

In this study, we employ the relativistic coupled-cluster method with single and double excitations, and perturbative triples [CCSD(T)].
Our results are in good agreement with the recommended values from Ref.~\citenum{Schwerdtfeger2019} and align with other theoretical predictions for Cu \cite{Neogrády1997, Maroulis2006, Mohr2009}, Ag \cite{gobre2016efficient, Dzuba2021, Tomza2021, Smialkowski2021}, and Au \cite{Neogrády1997, Wesendrup2000, Maroulis2006, Gould2016a, Tomza2021}.
We also provide a quantitative analysis of relativistic and electron-correlation corrections.
By employing the empirical error estimation method proposed in Ref.~\citenum{Cheng2024a}, we present final recommended values with significantly reduced uncertainties for Ag and Au, improving upon the values recommended in Ref.~\citenum{Schwerdtfeger2019}.

The structure of the remainder of this paper is as follows.
Section~\ref{sec:methods} introduces the computational methods, followed by computational details in Sec.~\ref{sec:details}.
Results are presented and discussed in Sec.~\ref{sec:results}.
A summary is given in Sec.~\ref{sec:summary}.
Atomic units are used throughout.

    \section{Methods}
    \label{sec:methods}
    The methods employed in this study are discussed in detail in Ref.~\citenum{Cheng2024a}, and the essential components are summarized here for completeness.

\subsection{Relativistic framework}
\label{subsec:relativistic-framework}
In this study, all calculations are performed within a fully relativistic framework using a four-component formalism.
The relativistic effects included in these calculations are divided into two components: spin-orbit coupling (SOC) effects and scalar-relativistic effects, which account for the contraction or expansion of radial electron densities due to relativistic corrections.\cite{Yu2015}

\subsubsection{Four-component Dirac-Coulomb Hamiltonian}
The standard relativistic electronic structure theory for four-component calculations is based on the Dirac-Coulomb-Breit (DCB) Hamiltonian \cite{Fleig2012}:
\begin{align}
    \hat{H}_\text{DCB} = \sum_i \hat{h}_\text{D}(i) + \sum_{i<j} \hat{g}_{ij} + \sum_{A<B} V_{AB},
    \label{eq:H_DCB}
\end{align}
where $i$ and $j$ label the electrons, $V_{AB}$ describes the nucleus-nucleus interactions, and $\hat{h}_\text{D}$ is the one-electron Dirac Hamiltonian.
Without the presence of external electric fields, $\hat{h}_\text{D}$ is expressed as
\begin{align}
    \hat{h}_\text{D}(i) = c \bm{\alpha}_i \cdot \bm{p}_i + c^2 \beta_i + \sum_A V_{iA},
    \label{eq:h_D}
\end{align}
where $c$ is the speed of light, $\bm{p}_i$ is the momentum operator, and $\bm{\alpha}$ and $\beta$ are the Dirac matrices given by:
\begin{align}
    \alpha_x &=
    \begin{pmatrix}
        0_2 & \sigma_x \\
        \sigma_x & 0_2
    \end{pmatrix},
    \quad
    \alpha_y =
    \begin{pmatrix}
        0_2 & \sigma_y \\
        \sigma_y & 0_2
    \end{pmatrix},
    \quad
    \nonumber \\
    \alpha_z &=
    \begin{pmatrix}
        0_2 & \sigma_z \\
        \sigma_z & 0_2
    \end{pmatrix},
    \quad
    \beta =
    \begin{pmatrix}
        I_2 & 0_2 \\
        0_2 & -I_2
    \end{pmatrix},
    \label{eq:alpha}
\end{align}
where $\sigma_x$, $\sigma_y$, and $\sigma_z$ are the Pauli spin matrices, and $0_2$ and $I_2$ are $2 \times 2$ zero and unit matrices, respectively.
The term $V_{iA}$ accounts for the electron-nucleus interaction between electron $i$ and nucleus $A$.

In the case of two-electron interactions, the relativistic corrections are included up to second order, represented by the Coulomb-Breit interaction in the Coulomb gauge \cite{Saue2011}:
\begin{align}
    \hat{g}_{ij} &= \hat{g}^\text{Coulomb} + \hat{g}^\text{Gaunt} + \hat{g}^\text{gauge}  \nonumber \\
                 &= \frac{1}{r_{ij}} - \frac{\bm{\alpha}_i \cdot \bm{\alpha}_j}{r_{ij}} + \frac{(\bm{\alpha}_i \cdot \bm{r}_{ij})(\bm{\alpha}_j \cdot \bm{r}_{ij})}{2r_{ij}^3},
    \label{eq:V_ij}
\end{align}
where the Gaunt and gauge terms collectively form the Breit interaction.
When only the Coulomb term is considered, the DCB Hamiltonian simplifies to the Dirac-Coulomb (DC) Hamiltonian:
\begin{align}
    \hat{H}_\text{DC} = \sum_i \hat{h}_\text{D}(i) + \sum_{i<j} \frac{1}{r_{ij}} + \sum_{A<B} V_{AB}.
    \label{eq:H_DC_approx}
\end{align}
In this study, we use the DC Hamiltonian for all calculations.
The Gaunt or Breit term is employed solely to estimate uncertainties.

\subsubsection{Exact two-component Dirac Hamiltonian}
Due to the high computational cost associated with four-component relativistic calculations, two-component approximations, such as the Douglas-Kroll-Hess (DKH) Hamiltonian \cite{Douglas1974, Hess1985, Hess1986} and the zeroth-order regular approximation (ZORA) \cite{Chang1986, vanLenthe1994, vanLenthe1996}, are frequently employed.
However, these methods are restricted to finite-order approximations.

An alternative infinite-order approach, the exact two-component (X2C) method, was implemented by Ilias and Saue \cite{Ilias2007}.
The X2C method builds on earlier work by Dyall \cite{Dyall1997}, with subsequent developments detailed in Refs.~\citenum{Dyall2002a,Kutzelnigg2005,Filatov2007,Peng2007,Ilias2007,Liu2009,Saue2011,Peng2012,Liu2014,Liu2016a,Liu2020}.

The X2C framework separates relativistic effects into scalar-relativistic (spin-free) and spin-orbit coupling (SOC) components \cite{Dyall2001}.
In this study, we employ the spin-free X2C method to compute nonrelativistic and scalar-relativistic properties, following the approaches described in Refs.~\citenum{Saue2011, Saue2020}.

\subsection{Relativistic and electron-correlation effects}
\label{subsec:electron-correlation}
The uncorrelated reference calculations are performed using the Dirac-Hartree-Fock (DHF) method, where the terms ``orbital'' and ``spinor'' are used to describe the electronic state without and with SOC effects, respectively.
Post-Hartree-Fock methods, including second-order M\o ller-Plesset perturbation theory (MP2) \cite{vanStralen2005a} and CC methods such as CCSD and CCSD(T) \cite{Visscher1996}, are used to account for electron-correlation effects.

The abbreviations NR-CC, SR-CC, and DC-CC refer to CC calculations under nonrelativistic (NR), scalar-relativistic (SR), and DC relativistic effects, respectively.
In correlated calculations, occupied and virtual orbitals or spinors are truncated to reduce computational costs.
Orbitals or spinors from DHF are divided into inner-core, outer-core, valence, and virtual orbitals, with only outer-core and valence orbitals being correlated.

The electron-correlation contributions to a property $X$ (polarizability or hyperpolarizability) are defined as
\begin{align}
    \Delta X_c^n = X_{\text{CCSD(T)}}^n - X_{\text{DHF}}^n,
    \label{eq:alpha_corr}
\end{align}
where $n$ represents either NR, SR and DC relativistic effects.

The scalar-relativistic effects on $X$ are defined as
\begin{align}
    \Delta X_r^\text{SR} = X_\text{CCSD(T)}^\text{SR} - X_\text{CCSD(T)}^\text{NR},
    \label{eq:rel_effects_sr}
\end{align}
where $X_\text{CCSD(T)}^\text{SR}$ and $X_\text{CCSD(T)}^\text{NR}$ denote the scalar-relativistic and nonrelativistic $X$, respectively.
The SOC contribution to $X$ is then given by
\begin{align}
    \Delta X_r^\text{SOC} = X_\text{CCSD(T)}^\text{DC} - X_\text{CCSD(T)}^\text{SR},
    \label{eq:rel_effects_so}
\end{align}
where $X_\text{CCSD(T)}^\text{DC}$ is the DC relativistic value of $X$, calculated using the CCSD(T) method.
It should be noted that the correlation level used in scalar-relativistic calculations might differ between Eqs.~\eqref{eq:rel_effects_sr} and~\eqref{eq:rel_effects_so}.
This is because DC calculations at the same correlation level are much more costly than their SR counterparts.
Consequently, when the correlation levels differ, SR$_n$ and SR$_d$ are used to denote these values, respectively.

The total relativistic correction to $X$ is the sum of scalar-relativistic and SOC contributions
\begin{align}
    \Delta X_r^\text{DC} = \Delta X_r^\text{SR} + \Delta X_r^\text{SOC}.
    \label{eq:rel_effects_dc}
\end{align}

\subsection{Finite-field methods}
\label{subsec:finite-field-methods}
Static dipole polarizabilities and hyperpolarizabilities are computed using the finite-field method \cite{Das1998}.
The energy of an atom in an external electric field of strength $F_z$ along the $z$ axis is given by
\begin{align}
    E(F_z) \approx E_0 - \frac{1}{2}\alpha F_z^2 - \frac{1}{4!} \gamma F_z^4,
    \label{eq:ff_d_2}
\end{align}
where $E_0$ is the field-free energy, $\alpha$ is the dipole polarizability, and $\gamma$ is the dipole hyperpolarizability.
Least-squares fitting is used to extract $\alpha$ and $\gamma$ from computed energies at different field strengths.
In cases where $\gamma$ results in unphysical values, only $\alpha$ is retained using
\begin{align}
    E(F_z) \approx E_0 - \frac{1}{2}\alpha F_z^2.
    \label{eq:ff_d_1}
\end{align}

In this work, Eq.~\eqref{eq:ff_d_1} is used exclusively for DC CCSD(T) calculations because negative values of $\gamma$ are observed for Cu, Ag, and Au.
It is important to note that Eq.~\eqref{eq:ff_d_1} can be considered a special case of Eq.~\eqref{eq:ff_d_2} with $\gamma = 0$, which typically results in a larger $\alpha$ than the accurate value.
A more appropriate approach is to assign an approximate positive value to $\gamma$, denoted as $\gamma_\text{approx.}$:
\begin{align}
    E(F_z) \approx E_0 - \frac{1}{2}\alpha F_z^2 - \frac{1}{4!} \gamma_\text{approx.} F_z^4.
    \label{eq:ff_d_1_fixed}
\end{align}
The choice of $\gamma_\text{approx.}$ is discussed in Sec.~\ref{sec:results}.

In addition, the standard errors of the regression coefficients are evaluated using the residuals from the least-square solutions \cite{Williams2016}, which has been implemented in Ref.~\citenum{pydirac}.
This error is treated as the numerical fitting error, which is discussed in Sec.~\ref{subsec:uncertainty}.

\subsection{Uncertainty estimation}
\label{subsec:uncertainty}
Uncertainties in computed polarizabilities are estimated following the composite scheme \cite{Kallay2011, Yu2015, Irikura2021, Cheng2024a}.
The total uncertainty is expressed as:
\begin{align}
    P_\text{final} = &P_\text{CCSD} + \Delta P_\text{basis} + \Delta P_\text{(T)} \nonumber \\
                     &+ \Delta P_\text{core} + \Delta P_\text{vir} + \Delta P_\text{fitting} + \Delta P_\text{SOC} + \Delta P_\text{others},
\end{align}
where $P_\text{final}$ is the final value, and the different $\Delta P$ terms represent contributions from basis set incompleteness, triple excitations, core-electron correlation, virtual orbital truncation, numerical fitting, SOC effects, and other effects, respectively.
These contributions are added in quadrature to estimate the total uncertainty.
The detailed expression of each term is described in Ref.~\citenum{Cheng2024a}.

    \section{Computational Details}
    \label{sec:details}
    In this work, we primarily use the uncontracted Dyall quadruple-$\zeta$ family basis sets \cite{Dyall2004, Dyall2007, Dyall2010}.
The original dyall.cv4z basis sets are augmented with additional functions, extending each type of function in an even-tempered manner.
The exponential coefficients for the augmented functions are determined using the equation $\zeta_{N+1} = \zeta_N^2 /\zeta_{N-1}$, where $\zeta_N$ and $\zeta_{N-1}$ are the smallest exponents for each atomic shell in the default basis sets \cite{Yu2015}.
These augmented basis sets are labeled as s-aug-dyall.cv4z for single augmentations and d-aug-dyall.cv4z for double augmentations.

In general, orbitals within an energy range of -20 to 25 a.u. are correlated in the CC calculations.
The convergence criterion for these calculations is set to $10^{-10}$.

Electric fields with strengths of 0.000, 0.0005, 0.001, 0.002, and 0.005 a.u. are applied to each element to calculate the dipole polarizabilities.
All calculations are performed using the \texttt{DIRAC18} package \cite{DIRAC18}.
The resulting energies are then fitted to Eqs.~\eqref{eq:ff_d_2}-\eqref{eq:ff_d_1_fixed} using a least-squares method to obtain the dipole polarizabilities.

Each calculation is uniquely identified by a combination of computational method, basis set, and correlation level, represented by a string such as ``2C-SR-CC@s-aug-ANO-RCC@(core 3)[vir 279]''.
The components of this identifier are separated by the delimiter ``@''.
The first component refers to the computational method, which can be NR-CC, SR-CC, or DC-CC.
The prefix ``2C'' or ``4C'' indicates whether the two-component or four-component relativistic Hamiltonian is used, respectively.
The second component denotes the basis set employed in the calculation.
The final component represents the correlation level, which determines the accuracy of the correlated calculations.
In this study, the correlation level is described by the number of active electrons and virtual orbitals, expressed as ``(core N)[vir M]'', where $N$ refers to the sum of outer-core and valence electrons, and $M$ denotes the number of virtual orbitals.

In the CC module \cite{DIRAC18}, various correlated methods such as DHF, MP2, CCSD, and CCSD(T) are employed, all of which follow the same identifier format.
The percentage error $\delta_m$ of a property $X=\alpha~\text{or}~\gamma$ is defined as:
\begin{align}
    \delta_m = \frac{X_m - X_\text{CCSD(T)}}{X_\text{CCSD(T)}} \times 100\%,
\end{align}
where $m$ represents the method used, such as DHF, MP2, or CCSD.
The CCSD(T) value is used as the reference, denoted by $X_\text{CCSD(T)}$.

    \section{Results}
    \label{sec:results}
    In this section, we present the computational values for group 11 elements and discuss the impact of relativistic effects and electron-correlation contributions on dipole polarizabilities $\alpha$.

Table~\ref{tab:dipole_group__11__2__0} lists all results for Cu, Ag, and Au obtained by fitting Eq.~\eqref{eq:ff_d_2}.
The corresponding results of hyperpolarizabilities $\gamma$ are provided in Table S1 of the Supporting Information.

{\scriptsize
\begin{center}
\scriptsize
\begin{longtable}{ccccccc}

                \caption{
                Dipole polarizability results ($\alpha$ in a.u.) for group 11 elements.
                The error bar represents the uncertainty due to the numerical fitting procedure
                ($\Delta P_\text{fitting}$) for errors greater than 0.005 a.u.
                }\\

\hline
\multicolumn{1}{c}{\textbf{Z}} &
\multicolumn{1}{c}{\textbf{Atom}} &
\multicolumn{1}{c}{\textbf{State}} &
\multicolumn{1}{c}{\textbf{$\alpha$ (a.u.)}} &
\multicolumn{1}{c}{\textbf{$\delta$ (\%)}} &
\multicolumn{1}{c}{\textbf{Method}} &
\multicolumn{1}{c}{\textbf{Comments}} \\ \hline
\endfirsthead

\multicolumn{7}{l}
{{\bfseries \tablename\ \thetable{}. continued.}} \\ \hline
\multicolumn{1}{c}{\textbf{Z}} &
\multicolumn{1}{c}{\textbf{Atom}} &
\multicolumn{1}{c}{\textbf{State}} &
\multicolumn{1}{c}{\textbf{$\alpha$ (a.u.)}} &
\multicolumn{1}{c}{\textbf{$\delta$}} &
\multicolumn{1}{c}{\textbf{Method}} &
\multicolumn{1}{c}{\textbf{Comments}} \\ \hline
\endhead
\hline \multicolumn{7}{r}{{(continued)}} \\
\endfoot
\endlastfoot
         29 & Cu  & $^2S, M_L=0$        & $77.18 $ & 53.75 & DHF     & 2C-NR-CC@s-aug-dyall.cv4z@(core 29)[vir 299]\\
          &      &                & $31.61 \pm 0.15 $    & -37.03 & MP2     &            \\
          &      &                & $53.51  $    & 6.59 & CCSD    &            \\
          &      &                & $50.20 \pm 0.01$  & $--$   & CCSD(T) &             \\
         \cline{3-7}
          &      & $^2S, M_L=0$        & $70.58 $ & 53.85 & DHF     & 2C-SR-CC@dyall.cv4z@(core 29)[vir 227]\\
          &      &                & $28.92  $    & -36.97 & MP2     &            \\
          &      &                & $48.95  $    & 6.70 & CCSD    &            \\
          &      &                & $45.87 $  & $--$   & CCSD(T) &             \\
         \cline{3-7}
          &      & $^2S, M_L=0$        & $70.77 $ & 52.14 & DHF     & 2C-SR-CC@s-aug-dyall.cv4z@(core 19)[vir 271]\\
          &      &                & $28.93  $    & -37.81 & MP2     &            \\
          &      &                & $49.40  $    & 6.21 & CCSD    &            \\
          &      &                & $46.52 $  & $--$   & CCSD(T) &             \\
         \cline{3-7}
          &      & $^2S, M_L=0$        & $70.77 $ & 51.98 & DHF     & 2C-SR-CC@s-aug-dyall.cv4z@(core 29)[vir 299]\\
          &      &                & $29.01  $    & -37.70 & MP2     &            \\
          &      &                & $49.52  $    & 6.35 & CCSD    &            \\
          &      &                & $46.57 $  & $--$   & CCSD(T) &             \\
         \cline{3-7}
          &      & $^2S, M_L=0$        & $70.77 $ & 52.22 & DHF     & 2C-SR-CC@d-aug-dyall.cv4z@(core 29)[vir 371]\\
          &      &                & $28.55 \pm 0.01 $    & -38.58 & MP2     &            \\
          &      &                & $49.52  $    & 6.52 & CCSD    &            \\
          &      &                & $46.49 $  & $--$   & CCSD(T) &             \\
          \cline{3-7}
          &      & $^2S_{1/2}$    & $70.49         $  & 50.64 & DHF     & 4C-DC-CC@dyall.cv4z@(core 19)[vir 227] \\
          &      &                & $36.85     \pm 1.86    $  & -21.25 & MP2     &             \\
          &      &                & $48.89       $  & 4.47 & CCSD    &             \\
          &      &                & $46.79 \pm 0.25$  & $--$  & CCSD(T) &             \\
          \cline{3-7}
          &      & $^2S_{1/2}$    & $70.68         $  & 44.87 & DHF     & 4C-DC-CC@s-aug-dyall.cv4z@(core 19)[vir 271] \\
          &      &                & $50.35     \pm 8.28    $  & 3.20 & MP2     &             \\
          &      &                & $49.36       $  & 1.17 & CCSD    &             \\
          &      &                & $48.79 \pm 0.85$  & $--$  & CCSD(T) &             \\
        \hline
         47 & Ag  & $^2S, M_L=0$        & $105.52 $ & 68.39 & DHF     & 2C-NR-CC@s-aug-dyall.cv4z@(core 29)[vir 355]\\
          &      &                & $31.60 \pm 0.01 $    & -49.58 & MP2     &            \\
          &      &                & $68.10  $    & 8.68 & CCSD    &            \\
          &      &                & $62.67 $  & $--$   & CCSD(T) &             \\
         \cline{3-7}
          &      & $^2S, M_L=0$        & $81.82 $ & 61.50 & DHF     & 2C-SR-CC@dyall.cv4z@(core 29)[vir 283]\\
          &      &                & $26.59  $    & -47.52 & MP2     &            \\
          &      &                & $54.58  $    & 7.72 & CCSD    &            \\
          &      &                & $50.67 $  & $--$   & CCSD(T) &             \\
         \cline{3-7}
          &      & $^2S, M_L=0$        & $81.91 $ & 61.25 & DHF     & 2C-SR-CC@s-aug-dyall.cv4z@(core 29)[vir 355]\\
          &      &                & $25.78  $    & -49.25 & MP2     &            \\
          &      &                & $54.73  $    & 7.74 & CCSD    &            \\
          &      &                & $50.80 $  & $--$   & CCSD(T) &             \\
         \cline{3-7}
          &      & $^2S, M_L=0$        & $81.91 $ & 60.10 & DHF     & 2C-SR-CC@s-aug-dyall.cv4z@(core 19)[vir 331]\\
          &      &                & $26.96  $    & -47.30 & MP2     &            \\
          &      &                & $54.99  $    & 7.49 & CCSD    &            \\
          &      &                & $51.16 $  & $--$   & CCSD(T) &             \\
         \cline{3-7}
          &      & $^2S, M_L=0$        & $81.91 $ & 61.59 & DHF     & 2C-SR-CC@d-aug-dyall.cv4z@(core 29)[vir 427]\\
          &      &                & $23.42 \pm 0.17 $    & -53.81 & MP2     &            \\
          &      &                & $54.73  $    & 7.97 & CCSD    &            \\
          &      &                & $50.69 \pm 0.02$  & $--$   & CCSD(T) &             \\
          \cline{3-7}
          &      & $^2S_{1/2}$    & $81.63         $  & 58.50 & DHF     & 4C-DC-CC@dyall.cv4z@(core 19)[vir 283] \\
          &      &                & $36.27     \pm 2.24    $  & -29.57 & MP2     &             \\
          &      &                & $54.83       $  & 6.47 & CCSD    &             \\
          &      &                & $51.50 \pm 0.16$  & $--$  & CCSD(T) &             \\
          \cline{3-7}
          &      & $^2S_{1/2}$    & $81.72         $  & 57.40 & DHF     & 4C-DC-CC@s-aug-dyall.cv4z@(core 29)[vir 355] \\
          &      &                & $48.73     \pm 10.97    $  & -6.14 & MP2     &             \\
          &      &                & $54.63       $  & 5.23 & CCSD    &             \\
          &      &                & $51.91 \pm 0.58$  & $--$  & CCSD(T) &             \\
        \hline
         79 & Au  & $^2S, M_L=0$        & $106.66 \pm 0.01$ & 70.65 & DHF     & 2C-NR-CC@s-aug-dyall.cv4z@(core 43)[vir 381]\\
          &      &                & $26.25 \pm 0.01 $    & -58.00 & MP2     &            \\
          &      &                & $68.33 \pm 0.01 $    & 9.32 & CCSD    &            \\
          &      &                & $62.50 \pm 0.01$  & $--$   & CCSD(T) &             \\
         \cline{3-7}
          &      & $^2S, M_L=0$        & $48.06 $ & 33.20 & DHF     & 2C-SR-CC@dyall.cv4z@(core 43)[vir 309]\\
          &      &                & $24.52  $    & -32.04 & MP2     &            \\
          &      &                & $37.41  $    & 3.67 & CCSD    &            \\
          &      &                & $36.08 $  & $--$   & CCSD(T) &             \\
         \cline{3-7}
          &      & $^2S, M_L=0$        & $48.27 $ & 32.01 & DHF     & 2C-SR-CC@s-aug-dyall.cv4z@(core 33)[vir 381]\\
          &      &                & $25.36 \pm 0.01 $    & -30.64 & MP2     &            \\
          &      &                & $37.86  $    & 3.55 & CCSD    &            \\
          &      &                & $36.56 $  & $--$   & CCSD(T) &             \\
         \cline{3-7}
          &      & $^2S, M_L=0$        & $48.27 $ & 32.50 & DHF     & 2C-SR-CC@s-aug-dyall.cv4z@(core 43)[vir 381]\\
          &      &                & $24.88 \pm 0.01 $    & -31.71 & MP2     &            \\
          &      &                & $37.70 \pm 0.01 $    & 3.49 & CCSD    &            \\
          &      &                & $36.43 \pm 0.01$  & $--$   & CCSD(T) &             \\
         \cline{3-7}
          &      & $^2S, M_L=0$        & $48.27 $ & 33.25 & DHF     & 2C-SR-CC@d-aug-dyall.cv4z@(core 43)[vir 479]\\
          &      &                & $22.66 \pm 1.68 $    & -37.46 & MP2     &            \\
          &      &                & $37.62 \pm 0.01 $    & 3.84 & CCSD    &            \\
          &      &                & $36.23 \pm 0.1$  & $--$   & CCSD(T) &             \\
          \cline{3-7}
          &      & $^2S_{1/2}$    & $47.82         $  & 25.57 & DHF     & 4C-DC-CC@dyall.cv4z@(core 33)[vir 309] \\
          &      &                & $62.96     \pm 23.53    $  & 65.33 & MP2     &             \\
          &      &                & $37.53    \pm 0.01   $  & -1.46 & CCSD    &             \\
          &      &                & $38.08 \pm 1.17$  & $--$  & CCSD(T) &             \\
          \cline{3-7}
          &      & $^2S_{1/2}$    & $48.05         $  & 29.42 & DHF     & 4C-DC-CC@s-aug-dyall.cv4z@(core 33)[vir 381] \\
          &      &                & $34.52     \pm 5.68    $  & -7.03 & MP2     &             \\
          &      &                & $37.85    \pm 0.01   $  & 1.93 & CCSD    &             \\
          &      &                & $37.13 \pm 0.35$  & $--$  & CCSD(T) &             \\
    \hline
    \label{tab:dipole_group__11__2__0}
\end{longtable}
\end{center}
 }

To ensure the reliability of the calculated dipole polarizabilities, a systematic convergence study is performed using different basis sets: dyall.cv4z, s-aug-dyall.cv4z, and d-aug-dyall.cv4z.
This analysis involves evaluating the scalar-relativistic DHF, CCSD, and CCSD(T) energies, as well as $\alpha_\text{CCSD(T)}^\text{SR}$, as a function of the basis set size.
The goal is to identify the optimal basis set that achieves a balance between computational efficiency and numerical accuracy for the elements Cu, Ag, and Au.
The convergence trends for these properties with respect to the basis set size are summarized in Table~\ref{tab:convergence}.

The results for $\alpha$ obtained at the SR CCSD(T)/s-aug-dyall.cv4z level exhibit convergence compared to those obtained with the dyall.cv4z and d-aug-dyall.cv4z basis sets for all elements.
Therefore, the s-aug-dyall.cv4z basis is employed as the most accurate method at each relativistic level, balancing computational cost and precision.

\begin{table}
    \scriptsize
    \caption{
        Convergence of scalar-relativistic DHF energy ($E_\text{DHF}$), CCSD energy ($E_\text{CCSD}$), CCSD(T) energy ($E_\text{CCSD(T)}$) at $F_z = 0.000$, and dipole polarizability ($\alpha_\text{CCSD(T)}^\text{SR}$) with respect to basis set size for Cu, Ag, and Au.
        All values are in a.u.
    }
    \label{tab:convergence}
    \begin{tabular}{cccccccc}
        \toprule
        Atom &    Basis            & Basis size & Orbitals &  $E_\text{DHF}$ & $E_\text{CCSD}$ &$E_\text{CCSD}$ & $\alpha_\text{CCSD(T)}^\text{SR}$ \\
        \midrule
        Cu   &dyall.cv4z       & 324 & 30s20p12d6f4g2h   & -1653.1918406590 & -1654.0535048327& -1654.0828444931 & 45.87 \\
             &s-aug-dyall.cv4z & 380 & 31s21p13d7f5g3h   & -1653.1918481156 & -1654.0544974630& -1654.0839493137 & 46.57 \\
             &d-aug-dyall.cv4z & 436 & 32s22p14d8f6g4h   & -1653.1918485020 & -1654.0546535699& -1654.0841101325 & 46.49 \\
        Ag   &dyall.cv4z       & 418 & 33s25p17d7f5g3h   & -5312.8798587798 & -5313.9286242230& -5313.9584356052 & 50.67 \\
             &s-aug-dyall.cv4z & 474 & 34s26p18d8f6g4h   & -5312.8798590356 & -5313.9288485239& -5313.9586754851 & 50.80 \\
             &d-aug-dyall.cv4z & 530 & 35s27p19d9f7g5h   & -5312.8798590826 & -5313.9288809884& -5313.9587087835 & 50.69 \\
        Au   &dyall.cv4z       & 585 & 34s30p19d13f7g4h1i& -19009.0260213679& -19010.6236646443& -19010.6636563938& 36.08 \\
             &s-aug-dyall.cv4z & 669 & 35s31p20d14f8g5h2i& -19009.0260400275& -19010.6243806840& -19010.6644004336& 36.43 \\
             &d-aug-dyall.cv4z & 753 & 36s32p21d15f9g6h3i& -19009.0260408236& -19010.6354715241& -19010.6761765631& 36.23 \\
        \bottomrule
    \end{tabular}
\end{table}

Table~\ref{tab:dipole_group_11} summarizes the most accurate results for $\alpha$ for group 11 elements and compares them with the recommended values from Ref.~\citenum{Schwerdtfeger2019}.
The corresponding results for $\gamma$ are presented in Table S2 in the Supporting Information.
The DC CCSD(T) results obtained from Eq.~\eqref{eq:ff_d_1} are also presented in Table~\ref{tab:dipole_group_11}, while other results from Eq.~\eqref{eq:ff_d_1} are found in Table S3 in the Supporting Information.
A summary of the most accurate results from Eq.~\eqref{eq:ff_d_1} is listed in Table S4.

\begin{table}
\centering
\caption{
            Static dipole polarizabilities (in a.u.) with nonrelativistic,
            scalar-relativistic, and full Dirac-Coulomb relativistic effects
            for group 11 elements.
            SR$_n$ represents the SR results obtained using the same correlation level as in the
            NR calculations, while SR$_d$ represents the SR results evaluated using the same
            correlation level as in the DC calculations.
            The error due to the numerical fitting procedure
            ($\Delta P_\text{fitting}$) is shown
            as the error bar. The recommended values (Rec.), including the uncertainty estimation
            as the error bars, are also listed and compared to the counterparts from
            Ref.~\citenum{Schwerdtfeger2019}.
            }
\label{tab:dipole_group_11}
\scriptsize \begin{tabular}{llllll}
\toprule
                                 &      &      &                Cu &                Ag &                 Au \\
$\hat{H}$ & State & Method &                   &                   &                    \\
\midrule
NR & $^2S$ & DHF &           $77.18$ &          $105.52$ &  $106.66 \pm 0.01$ \\
                                 &      & CCSD &           $53.51$ &           $68.10$ &   $68.33 \pm 0.01$ \\
                                 &      & CCSD(T) &  $50.20 \pm 0.01$ &           $62.67$ &   $62.50 \pm 0.01$ \\
SR$_n$ & $^2S$ & DHF &           $70.77$ &           $81.91$ &            $48.27$ \\
                                 &      & CCSD &           $49.52$ &           $54.73$ &            $37.70$ \\
                                 &      & CCSD(T) &           $46.57$ &           $50.80$ &   $36.43 \pm 0.01$ \\
SR$_d$ & $^2S$ & DHF &           $70.77$ &           $81.91$ &            $48.27$ \\
                                 &      & CCSD &           $49.40$ &           $54.73$ &            $37.86$ \\
                                 &      & CCSD(T) &           $46.52$ &           $50.80$ &            $36.56$ \\
DC & $^2S_{1/2}$ & DHF &           $70.68$ &           $81.72$ &            $48.05$ \\
                                 &      & CCSD &           $49.36$ &           $54.63$ &            $37.85$ \\
                                 &      & CCSD(T) &  $48.79 \pm 0.85$ &  $51.91 \pm 0.58$ &   $37.13 \pm 0.35$ \\
DC [from Eq.~\eqref{eq:ff_d_1}] & $^2S_{1/2}$ & CCSD(T) &  $47.01 \pm 0.17$ &  $51.13 \pm 0.09$ &   $36.70 \pm 0.05$ \\
DC [from Eq.~\eqref{eq:ff_d_1_fixed}] & $^2S_{1/2}$ & CCSD(T) &  $46.91 \pm 0.18$ &  $50.98 \pm 0.10$ &   $36.68 \pm 0.05$ \\
Rec. & $--$ & $--$ &  $46.91 \pm 1.31$ &  $50.98 \pm 1.93$ &   $36.68 \pm 0.78$ \\
Ref.~\citenum{Schwerdtfeger2019} & $--$ & $--$ &      $46.5\pm0.5$ &        $55.0\pm8$ &         $36.0\pm3$ \\
\bottomrule
\end{tabular}
\end{table}

All central values of the DC CCSD(T) $\alpha$ obtained by fitting Eq.~\eqref{eq:ff_d_2} are slightly higher than the results obtained from Eq.~\eqref{eq:ff_d_1}, as shown in Table~\ref{tab:dipole_group_11}.
The corresponding differences are 1.78, 0.78, and 0.37 a.u. for Cu, Ag, and Au, respectively.
However, the trend is reversed for $\alpha$ evaluated at the DC CCSD level, where the differences in central values are -0.09, -0.16, and -0.04 a.u. for Cu, Ag, and Au, respectively.
This discrepancy arises because $\gamma$ defined in Eq.~\eqref{eq:ff_d_2} is negative for Cu, Ag, and Au when DC CCSD(T) energies are used, as shown in Table S4.
The negative $\gamma$ likely results from perturbative treatment of triplet states in contrast to iterative treatment of singlets and doublets in CCSD.
In practice, $\gamma$ should be positive, implying that the central values of the DC results may be overestimated.
To address this, we applied an approximate positive $\gamma$ instead of zero, as in Eq.~\eqref{eq:ff_d_1}.
Several methods can be used to estimate $\gamma$ for group 11 elements.
First, one can use $\gamma$ from DC CCSD calculations, assuming a small contribution of $\Delta P_\text{(T)}$ to $\gamma$.
For NR and SR values, data obtained from Eq.~\eqref{eq:ff_d_2} are more accurate due to the reasonable $\gamma$ predicted.
Second, $\gamma$ from SR CCSD(T) calculations can be used, assuming a small SOC contribution to $\gamma$.
Finally, a composite scheme can be employed to obtain an approximate $\gamma$, as described in Ref.~\citenum{Yu2015}, where the final $\gamma$ is the sum of SR CCSD(T) $\gamma$ and the SOC contribution to $\gamma$ evaluated at the CCSD level.
In this study, we tested all three methods, and the differences between the latter two are less than 0.01 a.u. for all elements, while the difference between the first two is less than 0.08 a.u. for Cu and 0.03 a.u. for Ag and Au.
For simplicity, we use the results from the third method as the values from Eq.~\eqref{eq:ff_d_1_fixed}, listed in Table~\ref{tab:dipole_group_11}.

A significantly larger uncertainty due to $\Delta P_\text{fitting}$ is observed in the DC CCSD(T) values for Cu, Ag, and Au, whereas no such uncertainty is present in the DC CCSD values or in any NR and SR values.
Additionally, a similar trend is observed in the MP2 results for certain NR and SR values, as shown in Table \ref{tab:dipole_group__11__2__0}.
One possible explanation is that the orbitals are not necessarily semi-canonical, as assumed in the derivation of perturbative corrections, particularly in open-shell calculations.\cite{Visscher1996}
This observation is also consistent with previously reported DC CCSD(T) values for group 1 and group 13 elements.\cite{Cheng2024a}

\subsection{Comparison with literature}
\label{ssec:alpha_compare-liter-group-11}

The theoretical and experimental values for Cu, Ag, and Au are summarized in Refs.~\citenum{Schwerdtfeger2019, Schwerdtfeger2023}.
For consistency, these values are compiled in Table~\ref{tab:group_11_others}, sorted by publication year, to validate the results of this study.
Only computational and experimental values close to the recommended values in Ref.~\citenum{Schwerdtfeger2019} are considered in the following discussion.
For SR and NR results, only values computed using the CCSD(T) method are compared.
Additionally, for SR polarizability, only the $\text{SR}_n$ value from this work is considered when $\text{SR}_n$ and $\text{SR}_d$ differ, due to the smaller number of correlated electrons used in the latter calculations.
The deviation between the SR$_n$ and SR$_d$ values is found to be less than 0.4\%, as shown in Table~\ref{tab:dipole_group_11}.

{\scriptsize
    \begin{longtable}{lllllrl}
\caption{
    Summary of reference atomic dipole polarizabilities (in a.u.) for group 11 elements, as reported in Refs.~\citenum{Schwerdtfeger2019} and \citenum{Schwerdtfeger2023}.
    Data from Ref.~\citenum{Schwerdtfeger2019} are reproduced with permission. Copyright 2018 Taylor \& Francis.
    Data from Ref.~\citenum{Schwerdtfeger2023} are reproduced with permission. Copyright 2023 Peter Schwerdtfeger and Jeffrey K. Nagle.
    Comment definitions are provided in these references.
    }
\label{tab:group_11_others}\\
\toprule
 Z & Atom &                                                      Refs. &                 State &         $\alpha$ &  Year &                   Comments \\
\midrule
\endfirsthead

\toprule
 Z & Atom &                                                      Refs. &                 State &         $\alpha$ &  Year &                   Comments \\
\midrule
\endhead
\midrule
\multicolumn{7}{r}{{Continued on next page}} \\
\midrule
\endfoot

\bottomrule
\endlastfoot
29 &   Cu &                              [\citenum{Schwerdtfeger1994}] & $^2S_{1/2},3d^{{10}}$ &           $45.0$ &  1994 &            R, PP, QCISD(T) \\
   &      &                                [\citenum{Pou-Amérigo1995}] &       $^2S,3d^{{10}}$ &          $53.44$ &  1995 &                   NR, MCPF \\
   &      &                           [\citenum{Lide2004, Doolen1987}] & $^2S_{1/2},3d^{{10}}$ &      $41 \pm 10$ &  2004 &              R, Dirac, LDA \\
   &      &                                        [\citenum{Chu2004}] & $^2S_{1/2},3d^{{10}}$ &           $39.5$ &  2004 &                    SIC-DFT \\
   &      &                                       [\citenum{Roos2005}] & $^2S_{1/2},3d^{{10}}$ &   $40.7 \pm 4.1$ &  2005 &              R, DK, CASPT2 \\
   &      &                                      [\citenum{Kłos2005a}] & $^2S_{1/2},3d^{{10}}$ &   $43.7 \pm 4.4$ &  2005 &                R, DK, MRCI \\
   &      &                     [\citenum{Maroulis2006, Neogrády1997}] & $^2S_{1/2},3d^{{10}}$ & $46.50 \pm 0.35$ &  2006 &             R, DK, CCSD(T) \\
   &      &                                       [\citenum{Mohr2009}] & $^2S_{1/2},3d^{{10}}$ &          $46.98$ &  2009 &             R, DK, CCSD(T) \\
   &      &                        [\citenum{Mitroy2010a, Zhang2008a}] & $^2S_{1/2},3d^{{10}}$ &          $41.65$ &  2010 &                       CICP \\
   &      &                         [\citenum{Hohm2012, Sarkisov2006}] & $^2S_{1/2},3d^{{10}}$ &   $54.7 \pm 5.5$ &  2012 &                       exp. \\
   &      &                                         [\citenum{Ma2015}] & $^2S_{1/2},3d^{{10}}$ &   $58.7 \pm 4.7$ &  2015 &                       exp. \\
   &      &                                    [\citenum{Dyugaev2016}] &       $^2S,3d^{{10}}$ &           $51.8$ &  2016 &             semi-empirical \\
   &      &                                      [\citenum{Ernst2016}] & $^2S_{1/2},3d^{{10}}$ &   $42.6 \pm 4.3$ &  2016 &      DFT B3LYP/aug-cc-pVDZ \\
   &      &                                     [\citenum{Gould2016a}] & $^2S_{1/2},3d^{{10}}$ &           $41.7$ &  2016 &              TD-DFT (LEXX) \\
   &      &                                     [\citenum{Gould2016b}] & $^2S_{1/2},3d^{{10}}$ &           $46.1$ &  2016 &               TD-DFT (PGG) \\
   &      &                                     [\citenum{Gould2016b}] & $^2S_{1/2},3d^{{10}}$ &           $41.2$ &  2016 &              SIC-DFT (RXH) \\
   &      &                              [\citenum{Schwerdtfeger2019}] &                  $--$ &   $46.5 \pm 0.5$ &  2019 &                recommended \\
47 &   Ag &                                   [\citenum{Neogrády1997}] &  $^2S_{1/2}, 4d^{10}$ &   $55.3 \pm 0.5$ &  1997 &             R, DK, CCSD(T) \\
   &      &                                       [\citenum{Roos2005}] &        $^2S, 4d^{10}$ &           $36.7$ &  2005 &             R, DK, CCSD(T) \\
   &      &                     [\citenum{Maroulis2006, Neogrády1997}] &        $^2S, 4d^{10}$ & $52.46 \pm 0.52$ &  2006 &             R, DK, CCSD(T) \\
   &      &                                     [\citenum{Zhang2008a}] &  $^2S_{1/2}, 4d^{10}$ &          $46.17$ &  2008 &                       CICP \\
   &      &                    [\citenum{Schwerdtfeger1994, Mohr2009}] &        $^2S, 4d^{10}$ &           $52.2$ &  2009 &            R, PP, QCISD(T) \\
   &      &                                [\citenum{Bezchastnov2010}] &  $^2S_{1/2}, 4d^{10}$ &      $56 \pm 14$ &  2010 &                       exp. \\
   &      &                                       [\citenum{Hohm2012}] &  $^2S_{1/2}, 4d^{10}$ &   $63.1 \pm 6.3$ &  2012 &                       exp. \\
   &      &                                         [\citenum{Ma2015}] &  $^2S_{1/2}, 4d^{10}$ &   $45.9 \pm 7.4$ &  2015 &                       exp. \\
   &      &                                    [\citenum{Dyugaev2016}] &        $^2S, 4d^{10}$ &           $55.2$ &  2016 &             Semi-empirical \\
   &      &                                     [\citenum{Gould2016a}] &  $^2S_{1/2}, 4d^{10}$ &           $46.2$ &  2016 &              TD-DFT (LEXX) \\
   &      &                                     [\citenum{Gould2016b}] &  $^2S_{1/2}, 4d^{10}$ &           $63.3$ &  2016 &               TD-DFT (PGG) \\
   &      &                                     [\citenum{Gould2016b}] &  $^2S_{1/2}, 4d^{10}$ &           $57.3$ &  2016 &              SIC-DFT (RXH) \\
   &      &                             [\citenum{gobre2016efficient}] &  $^2S_{1/2}, 4d^{10}$ &          $50.60$ &  2016 &                    LR-CCSD \\
   &      &                                     [\citenum{A.Manz2019}] &  $^2S_{1/2}, 4d^{10}$ &             $55$ &  2019 &                  ECP, CCSD \\
   &      &                              [\citenum{Schwerdtfeger2019}] &                  $--$ &       $55 \pm 8$ &  2019 &                recommended \\
   &      &                     [\citenum{Tomza2021, Smialkowski2021}] &  $^2S_{1/2}, 4d^{10}$ &           $50.2$ &  2021 &           SR, ECP, CCSD(T) \\
   &      &                                      [\citenum{Dzuba2021}] &  $^2S_{1/2}, 4d^{10}$ &           $50.6$ &  2021 &                 R, CI+MBPT \\
   &      &                [\citenum{Lide2004, Bromley2002b}]          &  $^2S_{1/2}, 4d^{10}$ &           $48.4$ &  2023 &              R, Dirac, LDA \\
79 &   Au &                                  [\citenum{Henderson1997}] &  $^2S_{1/2}, 5d^{10}$ &       $30 \pm 4$ &  1997 &       R, HFR, HS, CI, CACP \\
   &      &                                  [\citenum{Wesendrup2000}] &        $^2S, 5d^{10}$ &           $34.9$ &  2000 &             R, DK, CCSD(T) \\
   &      &                                       [\citenum{Roos2005}] &  $^2S_{1/2}, 5d^{10}$ &   $39.1 \pm 9.8$ &  2005 &                       exp. \\
   &      &                     [\citenum{Maroulis2006, Neogrády1997}] &        $^2S, 5d^{10}$ & $36.06 \pm 0.54$ &  2006 &             R, DK, CCSD(T) \\
   &      & [\citenum{Schwerdtfeger1994, Mohr2009, Schwerdtfeger2000}] &        $^2S, 5d^{10}$ &           $35.1$ &  2009 &            R, PP, QCISD(T) \\
   &      &                             [\citenum{Hohm2012, Roos2005}] &        $^2S, 5d^{10}$ &   $27.9 \pm 4.2$ &  2012 &              R, DK, CASPT2 \\
   &      &                         [\citenum{Hohm2012, Sarkisov2006}] &  $^2S_{1/2}, 5d^{10}$ &   $49.1 \pm 4.9$ &  2012 &                       exp. \\
   &      &                                     [\citenum{Gould2016a}] &  $^2S_{1/2}, 5d^{10}$ &           $45.4$ &  2016 &              TD-DFT (LEXX) \\
   &      &                             [\citenum{gobre2016efficient}] &  $^2S_{1/2}, 5d^{10}$ &          $36.50$ &  2016 &                    LR-CCSD \\
   &      &                                     [\citenum{A.Manz2019}] &  $^2S_{1/2}, 5d^{10}$ &          $39.56$ &  2019 &                  ECP, CCSD \\
   &      &                              [\citenum{Schwerdtfeger2019}] &                  $--$ &       $36 \pm 3$ &  2019 &                recommended \\
   &      &                                      [\citenum{Dzuba2021}] &  $^2S_{1/2}, 5d^{10}$ &           $34.0$ &  2021 &                 R, CI+MBPT \\
   &      &                                      [\citenum{Tomza2021}] &  $^2S_{1/2}, 5d^{10}$ &           $36.3$ &  2021 &           SR, ECP, CCSD(T) \\
   &      &                               [\citenum{Centoducatte2022}] &  $^2S_{1/2}, 5d^{10}$ &           $34.2$ &  2022 &      R (ZORA), DFT (B3LYP) \\
   &      &                                   [\citenum{Sarkisov2022}] &  $^2S_{1/2}, 5d^{10}$ &       $40 \pm 8$ &  2022 &                       exp. \\
   &      &                                       [\citenum{Neto2023}] &  $^2S_{1/2}, 5d^{10}$ &           $34.1$ &  2023 & R (ATZP-ZORA), DFT (B3LYP) \\
\end{longtable}
 }

For Cu, the SR value (46.57) obtained in this study agrees well with previous theoretical results ($46.50 \pm 0.35$ \cite{Maroulis2006, Neogrády1997} and 46.98 \cite{Mohr2009}) obtained by the Douglas-Kroll (DK) CCSD(T) method.
The central value of the DC CCSD(T) result (47.01) is slightly higher than the recommended value (46.5) from Ref.~\citenum{Schwerdtfeger2019}, which is based primarily on calculations considering SR effects as proposed in Refs.~\citenum{Maroulis2006, Neogrády1997, Mohr2009}, as shown in Table~\ref{tab:group_11_others}.
This difference likely arises from the SOC contribution included in this study.
Specifically, the SOC contribution evaluated at the CCSD(T) level (0.49 a.u.) is much larger than the counterpart (-0.04/-0.02 a.u.) obtained at the DHF/CCSD level, which is negligible, as shown in Table~\ref{tab:dipole_group_11}.
However, this contribution is likely overestimated since $\gamma$ is set to zero in Eq.~\eqref{eq:ff_d_1}.
This is confirmed by the DC CCSD(T) value ($46.91 \pm 0.18$) obtained from Eq.~\eqref{eq:ff_d_1_fixed}, where a more reasonable positive $\gamma$ is used, and the SOC contribution to $\alpha$ is reduced to $\recCuSOC$, which is considered the primary source of error due to $\Delta P_\text{others}$.
It is worth noting that this error could be further reduced by comparing results with triples included iteratively in CC, also known as CCSDT.
However, the computational cost makes this approach impractical for this work.
Both the SR and DC results are lower than the lower bound of experimental values ($54.7 \pm 5.5$ \cite{Hohm2012, Sarkisov2006} and $58.7 \pm 4.7$ \cite{Ma2015}), suggesting the need for more precise experiments.

For Ag, the DC values are $51.91 \pm 0.58$, $51.13 \pm 0.09$, and $50.98 \pm 0.10$ from Eqs.~\eqref{eq:ff_d_2} to \eqref{eq:ff_d_1_fixed}, respectively.
The difference between the values from Eq.~\eqref{eq:ff_d_1} and Eq.~\eqref{eq:ff_d_1_fixed} is 0.15 a.u., reflecting the correction of $\gamma$.
All DC values align well with the recommended value ($55.0 \pm 8$) from Ref.~\citenum{Schwerdtfeger2019}.
The central value (50.98) from Eq.~\eqref{eq:ff_d_1_fixed} is slightly higher than the value (50.6) obtained using relativistic CI+MBPT \cite{Dzuba2021}.
This difference may result from higher-order electron-correlation treatment in this study.
Additionally, our SR value (50.80) agrees well with the result (50.60) from CC with the linear-response theory \cite{gobre2016efficient} and the recent value (50.2) obtained using SR CCSD(T) with the effective core potential (ECP) method \cite{Tomza2021, Smialkowski2021}.
While the DC central value falls within the uncertainty range of experimental values ($56 \pm 14$ \cite{Bezchastnov2010} and $45.9 \pm 7.4$ \cite{Ma2015}), the large uncertainty in these experimental values warrants improvement.

For Au, the DC values are $37.13 \pm 0.35$, $36.70 \pm 0.05$, and $36.68 \pm 0.05$ from Eqs.~\eqref{eq:ff_d_2} to \eqref{eq:ff_d_1_fixed}, respectively.
The difference between the values from Eq.~\eqref{eq:ff_d_1} and Eq.~\eqref{eq:ff_d_1_fixed} is 0.02 a.u., suggesting a negligible effect of $\gamma$.
All DC values agree well with the recommended value ($36.0 \pm 3$) from Ref.~\citenum{Schwerdtfeger2019}.
The SR CCSD(T) value ($36.43 \pm 0.01$) also agrees well with DK CCSD(T) values (34.9 \cite{Wesendrup2000} and $36.06 \pm 0.54$ \cite{Maroulis2006, Neogrády1997}) and the value (36.3) obtained by SR CCSD(T) with the ECP method \cite{Tomza2021}.
The SR CCSD value ($37.70$), as shown in Table~\ref{tab:dipole_group_11}, is also close to the result (36.50) from CCSD with the linear-response theory \cite{Gould2016a}.
The central values of all DC and SR CCSD(T) results fall within the uncertainty range of experimental values ($39.1 \pm 9.8$ \cite{Roos2005} and $40 \pm 8$ \cite{Sarkisov2022}).

\subsection{Uncertainty estimation}
\label{ssec:reliable-group-11}

The uncertainty excluding $\Delta P_\text{fitting}$ is estimated for Cu, Ag, and Au.
In general, the errors due to $\Delta P_\text{core}$ and $\Delta P_\text{vir}$ are approximated by the difference between $\text{SR}_n$ and $\text{SR}_d$.
Half of the difference between the values evaluated using the s-aug-dyall.cv4z and d-aug-dyall.cv4z basis sets is taken as the error due to the finite basis set, i.e., $\Delta P_\text{basis}$, for each atom.
The effect of $\Delta T$ in DC calculations is $\recCuErrorDeltaT$, $\recAgErrorDeltaT$, and $\recAuErrorDeltaT$, corresponding to the difference between the DC CCSD value (from Eq.~\eqref{eq:ff_d_2}) and the CCSD(T) value (from Eq.~\eqref{eq:ff_d_1_fixed}).
In this work, half of the $\Delta T$ value is used to estimate the error due to $\Delta P_\text{(T)}$.
The SOC effect, defined as the difference between the SR CCSD(T) result from Eq.~\eqref{eq:ff_d_2} and the DC CCSD(T) result from Eq.~\eqref{eq:ff_d_1_fixed}, i.e., $\recCuSOC$ for Cu, $\recAgSOC$ for Ag, and $\recAuSOC$ for Au, is used as the error due to $\Delta P_\text{others}$.

The errors arising from the contributions of Breit and QED effects are evaluated as well.
Specifically, an empirical relationship between the first ionization potential, denoted by $\text{IP}$ (in eV), and the atomic dipole polarizability (in a.u.) is given by Ref.~\citenum{Schwerdtfeger2019}:
\begin{align}
    \alpha = a \times \text{IP}^b,
\end{align}
where $a$ and $b$ are unknown empirical coefficients.
The coefficient $b$ can be determined, for example, by fitting the values from DC CCSD and CCSD(T) calculations:
\begin{align}
    b = \log \left(\frac{\alpha_\text{CCSD}^\text{DC}}{\alpha_\text{CCSD(T)}^\text{DC}}\right) \times
    \left[\log \left(\frac{\text{IP}_\text{CCSD}^\text{DC}}{\text{IP}_\text{CCSD(T)}^\text{DC}}\right) \right]^{-1},
    \label{eq:b}
\end{align}
where the subscript $\text{DC}$ indicates that the $\text{IP}$ values are computed at the DC level.

Based on this empirical relationship, the atomic dipole polarizability including Breit+QED effects can be obtained using the ionization potential with and without Breit+QED effects.
For instance, starting from $\alpha_\text{CCSD(T)}^\text{DC}$, we have:
\begin{align}
    \alpha_\text{CCSD(T)+Breit+QED}^\text{DC} = \alpha_\text{CCSD(T)}^\text{DC} \times \left(\frac{\text{IP}_\text{CCSD(T)+Breit+QED}^\text{DC}}{\text{IP}_\text{CCSD(T)}^\text{DC}} \right)^b,
\end{align}
where $b$ is obtained from Eq.~\eqref{eq:b}.

Next, the uncertainty $\Delta P_\text{others} \approx \Delta P_\text{Breit+QED}$, arising from the contribution of Breit and QED effects to $\alpha$, can be empirically evaluated as:
\begin{align}
    \Delta P_\text{Breit+QED} = \alpha_\text{CCSD(T)+Breit+QED}^\text{DC} - \alpha_\text{CCSD(T)}^\text{DC}.
\end{align}

The values of $\text{IP}^\text{DC}_\text{CCSD}$, $\text{IP}^\text{DC}_\text{CCSD(T)}$, and $\text{IP}^\text{DC}_\text{CCSD(T)+Breit+QED}$, taken from Ref.~\citenum{Pašteka2017}, are 9.1164, 9.2938, and 9.2288 eV, respectively.
The contribution of Breit+QED effects to $\alpha$ for Au is 0.42 a.u.

For Cu and Ag, only scalar-relativistic CCSD and CCSD(T) data are available.
The $\text{IP}$ data are taken from Ref.~\citenum{Neogrády1997}, obtained using DK CCSD and CCSD(T).
The $\text{IP}_\text{CCSD}^\text{DK}$ and $\text{IP}_\text{CCSD(T)}^\text{DK}$ are 7.584 and 7.733 eV for Cu, respectively, and 7.360 and 7.461 eV for Ag, respectively.
The empirical coefficient $b$ is still obtained using Eq.~\eqref{eq:b} due to the small SOC contribution to $\alpha$, but $\alpha_\text{CCSD}^\text{SR}$ and $\alpha_\text{CCSD(T)}^\text{SR}$ are used instead of DC values.
The Breit+QED effects on $\text{IP}$, evaluated at the DHF level, can be found in Ref.~\citenum{Thierfelder2010}.
Consequently, the contributions of Breit+QED effects to $\alpha$ for Cu and Ag are 0.13 a.u. and 0.44 a.u., respectively.

Moreover, $\text{IP}_\text{CCSD}^\text{DK}$ and $\text{IP}_\text{CCSD(T)}^\text{DK}$ are also available for Au in Ref.~\citenum{Neogrády1997}, and the Breit+QED effects on $\text{IP}$ can be found in Ref.~\citenum{Thierfelder2010}.
Using scalar-relativistic data, the contribution of Breit+QED effects to $\alpha$ is 0.47 a.u. for Au.
To be conservative, we take 0.47 a.u. as the uncertainty due to Breit+QED effects on $\alpha$.

In conclusion, the total uncertainties (excluding $\Delta P_\text{fitting}$) are $\recCuErrorNoFit$, $\recAgErrorNoFit$, and $\recAuErrorNoFit$ a.u. for Cu, Ag, and Au, respectively.
The final recommended values (Rec.), obtained from the most accurate calculations with the total uncertainty, including the corresponding $\Delta P_\text{fitting}$, are listed in Table~\ref{tab:dipole_group_11}, with values of $\recCu$, $\recAg$, and $\recAu$ a.u. for Cu, Ag, and Au, respectively.

\subsection{Correlation and relativistic effects on polarizabilities}
\label{ssec:corr_rel}

\begin{figure*}[h]
\centering
\includegraphics{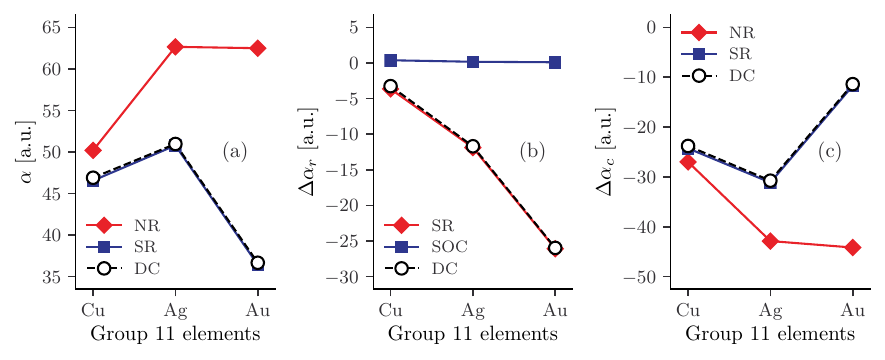}
\caption{
\label{fig:group_11}
    Polarizabilities (in a.u.) of group 11 elements.
    (a) Comparison of nonrelativistic, scalar-relativistic, and full-relativistic Dirac-Coulomb dipole polarizabilities $\alpha$.
    (b) Illustration of the influence of relativistic effects, including SR, SOC, and DC, on dipole polarizabilities $\Delta \alpha_r$.
    (c) Examination of the impact of electron-correlation effects on dipole polarizabilities $\Delta \alpha_c$ in the presence of various relativistic effects.
}
\end{figure*}

All quantities are computed using data from Table~\ref{tab:dipole_group_11}, and DC values obtained by fitting Eq.~\eqref{eq:ff_d_1_fixed} are used.
Figure~\ref{fig:group_11}(a) illustrates the relationship between dipole polarizabilities and atomic numbers for group 11 elements.
The nonrelativistic dipole polarizabilities for Ag and Au are similar, both exceeding that of Cu by more than 12 a.u.
In contrast, SR calculations show that Ag has the highest value, surpassing that of Au by more than 14 a.u.
The trend observed in the DC results follows that of the SR calculations due to minimal contribution from the SOC effect for group 11 elements.
This observation is further supported by Fig.~\ref{fig:group_11}(b), where SR, SOC, and DC relativistic contributions to $\alpha$ are evaluated at the CCSD(T) level.

Determining which system exhibits more significant SOC effects on dipole polarizabilities is challenging, as the SOC contributions are small: $\recCuSOC$, $\recAgSOC$, and $\recAuSOC$ for Cu, Ag, and Au, respectively, as evaluated at the CCSD(T) level.
At the CCSD level, these values, derived from Eq.~\eqref{eq:ff_d_2}, are -0.04, -0.10, and -0.01 for Cu, Ag, and Au, respectively.
However, at the DHF level, these values, also derived from Eq.~\eqref{eq:ff_d_2}, are -0.02, -0.19, and -0.22 for Cu, Ag, and Au, respectively, as shown in Table~\ref{tab:dipole_group_11}.

Electron-correlation effects on polarizabilities are explored in Fig.~\ref{fig:group_11}(c).
At the nonrelativistic level, the electron-correlation contribution increases in absolute value with increasing atomic number, with the difference between Ag and Cu being negligible.
However, both SR and DC calculations reveal the same trend: the maximum contribution is found for Ag, while the minimum is observed for Au.
Furthermore, the effect of $\Delta P_\text{(T)}$ in DC calculations, as discussed in Sec.~\ref{ssec:reliable-group-11}, is $\recCuErrorDeltaT$, $\recAgErrorDeltaT$, and $\recAuErrorDeltaT$ for Cu, Ag, and Au, respectively, which is approximately nine times smaller than the total correlation, defined as the difference between the DHF and CCSD(T) values, as shown in Fig.~\ref{fig:group_11}(c).
This supports the conclusion that using half of $\Delta P_\text{(T)}$ as the error estimate for higher-order correlation effects is appropriate.

    \section{Conclusion}
    \label{sec:summary}
    In this study, we have calculated the static dipole polarizabilities of group 11 elements using the finite-field method combined with relativistic CCSD(T) calculations.
Our results showed good agreement with the recommended values from the literature.
The final recommended dipole polarizability values, with associated uncertainties, are $\recCu$ a.u. for Cu, $\recAg$ a.u. for Ag, and $\recAu$ a.u. for Au.
We also provided a systematic analysis of the impact of various relativistic effects on atomic dipole polarizabilities, including scalar-relativistic effects, SOC, and Dirac-Coulomb relativistic contributions.
The analysis indicated that scalar-relativistic effects are the dominant relativistic contribution for Cu, Ag, and Au, while SOC effects were generally negligible.
Finally, our investigation of electron correlation, in conjunction with relativistic effects, underscored its critical role in accurately determining the dipole polarizabilities of group 11 elements.

    \begin{acknowledgement}
            Y.C. acknowledges the Foundation of Scientific Research, Flanders (Grant No. G0A9717N) and the Research Board of Ghent University for their financial support.
    The resources and services used in this work were provided by the Flemish Supercomputer Center, funded by the Research Foundation, Flanders and the Flemish Government.
     \end{acknowledgement}

    \begin{suppinfo}
        The Supplementary Material includes a PDF document containing the atomic dipole hyperpolarizabilities obtained by fitting Eq.~\eqref{eq:ff_d_2} and the dipole polarizabilities obtained by fitting Eq.~\eqref{eq:ff_d_1} for Cu, Ag, and Au.
     \end{suppinfo}

    \begin{figure*}[h]
        \includegraphics{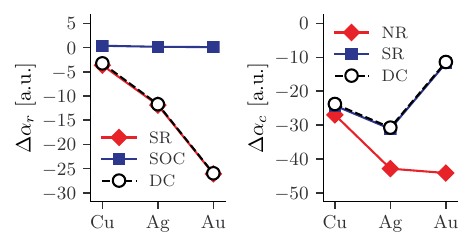}
        \caption{TOC}
    \end{figure*}

    \providecommand{\latin}[1]{#1}
\makeatletter
\providecommand{\doi}
  {\begingroup\let\do\@makeother\dospecials
  \catcode`\{=1 \catcode`\}=2 \doi@aux}
\providecommand{\doi@aux}[1]{\endgroup\texttt{#1}}
\makeatother
\providecommand*\mcitethebibliography{\thebibliography}
\csname @ifundefined\endcsname{endmcitethebibliography}
  {\let\endmcitethebibliography\endthebibliography}{}


\begin{mcitethebibliography}{101}
\providecommand*\natexlab[1]{#1}
\providecommand*\mciteSetBstSublistMode[1]{}
\providecommand*\mciteSetBstMaxWidthForm[2]{}
\providecommand*\mciteBstWouldAddEndPuncttrue
  {\def\EndOfBibitem{\unskip.}}
\providecommand*\mciteBstWouldAddEndPunctfalse
  {\let\EndOfBibitem\relax}
\providecommand*\mciteSetBstMidEndSepPunct[3]{}
\providecommand*\mciteSetBstSublistLabelBeginEnd[3]{}
\providecommand*\EndOfBibitem{}
\mciteSetBstSublistMode{f}
\mciteSetBstMaxWidthForm{subitem}{(\alph{mcitesubitemcount})}
\mciteSetBstSublistLabelBeginEnd
  {\mcitemaxwidthsubitemform\space}
  {\relax}
  {\relax}

\bibitem[Schwerdtfeger and Nagle(2019)Schwerdtfeger, and
  Nagle]{Schwerdtfeger2019}
Schwerdtfeger,~P.; Nagle,~J.~K. 2018 {{Table}} of Static Dipole
  Polarizabilities of the Neutral Elements in the Periodic Table. \emph{Mol.
  Phys.} \textbf{2019}, \emph{117}, 1200--1225\relax
\mciteBstWouldAddEndPuncttrue
\mciteSetBstMidEndSepPunct{\mcitedefaultmidpunct}
{\mcitedefaultendpunct}{\mcitedefaultseppunct}\relax
\EndOfBibitem
\bibitem[Bast \latin{et~al.}(2008)Bast, He{\ss}elmann, Sa{\l}ek, Helgaker, and
  Saue]{Bast2008}
Bast,~R.; He{\ss}elmann,~A.; Sa{\l}ek,~P.; Helgaker,~T.; Saue,~T. Static and
  Frequency-dependent Dipole--Dipole Polarizabilities of All Closed-shell Atoms
  up to Radium: A Four-component Relativistic {{DFT}} Study.
  \emph{ChemPhysChem} \textbf{2008}, \emph{9}, 445--453\relax
\mciteBstWouldAddEndPuncttrue
\mciteSetBstMidEndSepPunct{\mcitedefaultmidpunct}
{\mcitedefaultendpunct}{\mcitedefaultseppunct}\relax
\EndOfBibitem
\bibitem[Dyall(2011)]{Dyall2011}
Dyall,~K.~G. Relativistic Double-Zeta, Triple-Zeta, and Quadruple-Zeta Basis
  Sets for the 6d Elements {{Rf}}\textendash{{Cn}}. \emph{Theor. Chem. Acc.}
  \textbf{2011}, \emph{129}, 603--613\relax
\mciteBstWouldAddEndPuncttrue
\mciteSetBstMidEndSepPunct{\mcitedefaultmidpunct}
{\mcitedefaultendpunct}{\mcitedefaultseppunct}\relax
\EndOfBibitem
\bibitem[Dyall(2016)]{Dyall2016}
Dyall,~K.~G. Relativistic Double-Zeta, Triple-Zeta, and Quadruple-Zeta Basis
  Sets for the Light Elements {{H}}\textendash{{Ar}}. \emph{Theor. Chem. Acc.}
  \textbf{2016}, \emph{135}, 128\relax
\mciteBstWouldAddEndPuncttrue
\mciteSetBstMidEndSepPunct{\mcitedefaultmidpunct}
{\mcitedefaultendpunct}{\mcitedefaultseppunct}\relax
\EndOfBibitem
\bibitem[Dyall \latin{et~al.}(2023)Dyall, Tecmer, and Sunaga]{Dyall2023}
Dyall,~K.~G.; Tecmer,~P.; Sunaga,~A. Diffuse {Basis} {Functions} for
  {Relativistic} s and d {Block} {Gaussian} {Basis} {Sets}. \emph{J. Chem.
  Theory Comput.} \textbf{2023}, \emph{19}, 198--210\relax
\mciteBstWouldAddEndPuncttrue
\mciteSetBstMidEndSepPunct{\mcitedefaultmidpunct}
{\mcitedefaultendpunct}{\mcitedefaultseppunct}\relax
\EndOfBibitem
\bibitem[Ferreira \latin{et~al.}(2020)Ferreira, Campos, and
  Jorge]{Ferreira2020}
Ferreira,~I.~B.; Campos,~C.~T.; Jorge,~F.~E. All-Electron Basis Sets Augmented
  with Diffuse Functions for {{He}}, {{Ca}}, {{Sr}}, {{Ba}}, and Lanthanides:
  Application in Calculations of Atomic and Molecular Properties. \emph{J Mol
  Model} \textbf{2020}, \emph{26}, 95\relax
\mciteBstWouldAddEndPuncttrue
\mciteSetBstMidEndSepPunct{\mcitedefaultmidpunct}
{\mcitedefaultendpunct}{\mcitedefaultseppunct}\relax
\EndOfBibitem
\bibitem[Canal~Neto \latin{et~al.}(2021)Canal~Neto, Ferreira, Jorge, and {de
  Oliveira}]{CanalNeto2021}
Canal~Neto,~A.; Ferreira,~I.~B.; Jorge,~F.~E.; {de Oliveira},~A.~Z.
  All-Electron Triple Zeta Basis Sets for {{ZORA}} Calculations:
  {{Application}} in Studies of Atoms and Molecules. \emph{Chem. Phys. Lett.}
  \textbf{2021}, \emph{771}, 138548\relax
\mciteBstWouldAddEndPuncttrue
\mciteSetBstMidEndSepPunct{\mcitedefaultmidpunct}
{\mcitedefaultendpunct}{\mcitedefaultseppunct}\relax
\EndOfBibitem
\bibitem[Neto \latin{et~al.}(2021)Neto, {de Oliveira}, Jorge, and
  Camiletti]{Neto2021}
Neto,~A.~C.; {de Oliveira},~A.~Z.; Jorge,~F.~E.; Camiletti,~G.~G. {{ZORA}}
  All-Electron Double Zeta Basis Sets for the Elements from {{H}} to {{Xe}}:
  Application in Atomic and Molecular Property Calculations. \emph{J Mol Model}
  \textbf{2021}, \emph{27}, 1--9\relax
\mciteBstWouldAddEndPuncttrue
\mciteSetBstMidEndSepPunct{\mcitedefaultmidpunct}
{\mcitedefaultendpunct}{\mcitedefaultseppunct}\relax
\EndOfBibitem
\bibitem[Centoducatte \latin{et~al.}(2022)Centoducatte, {de Oliveira}, Jorge,
  and Camiletti]{Centoducatte2022}
Centoducatte,~R.; {de Oliveira},~A.~Z.; Jorge,~F.~E.; Camiletti,~G.~G. {{ZORA}}
  Double Zeta Basis Sets for Fifth Row Elements: {{Application}} in Studies of
  Electronic Structures of Atoms and Molecules. \emph{Comput. Theor. Chem.}
  \textbf{2022}, \emph{1207}, 113511\relax
\mciteBstWouldAddEndPuncttrue
\mciteSetBstMidEndSepPunct{\mcitedefaultmidpunct}
{\mcitedefaultendpunct}{\mcitedefaultseppunct}\relax
\EndOfBibitem
\bibitem[Neto \latin{et~al.}(2023)Neto, Jorge, and da~Cruz]{Neto2023}
Neto,~A.~C.; Jorge,~F.~E.; da~Cruz,~H. R.~C. All-Electron {{ZORA}} Triple Zeta
  Basis Sets for the Elements {{Cs}}--{{La}} and {{Hf}}--{{Rn}}. \emph{Chinese
  Phys. B} \textbf{2023}, \emph{32}, 093101\relax
\mciteBstWouldAddEndPuncttrue
\mciteSetBstMidEndSepPunct{\mcitedefaultmidpunct}
{\mcitedefaultendpunct}{\mcitedefaultseppunct}\relax
\EndOfBibitem
\bibitem[Gomes \latin{et~al.}(2024)Gomes, Jorge, and Neto]{Gomes2024}
Gomes,~C.~S.; Jorge,~F.~E.; Neto,~A.~C. All-electron basis sets for {H} to {Xe}
  specific for {ZORA} calculations: {Applications} in atoms and molecules.
  \emph{Chin. Phys. B} \textbf{2024}, \emph{33}, 083101\relax
\mciteBstWouldAddEndPuncttrue
\mciteSetBstMidEndSepPunct{\mcitedefaultmidpunct}
{\mcitedefaultendpunct}{\mcitedefaultseppunct}\relax
\EndOfBibitem
\bibitem[Sampaio \latin{et~al.}(2024)Sampaio, Jorge, and Neto]{Sampaio2024}
Sampaio,~G. R.~C.; Jorge,~F.~E.; Neto,~A.~C. {ZORA} {Basis} {Sets} of 5 and 6
  {Zeta} {Valence} {Qualities} for {H}-{Ar}: {Application} in {Calculations} of
  {Atomic} and {Molecular} {Properties}. \emph{Braz. J. Phys.} \textbf{2024},
  \emph{54}, 94\relax
\mciteBstWouldAddEndPuncttrue
\mciteSetBstMidEndSepPunct{\mcitedefaultmidpunct}
{\mcitedefaultendpunct}{\mcitedefaultseppunct}\relax
\EndOfBibitem
\bibitem[Newell and Tiesinga(2019)Newell, and Tiesinga]{Newell2019}
Newell,~D.; Tiesinga,~E. The {International} {System} of {Units} ({SI}), 2019
  {Edition}. \emph{NIST} \textbf{2019}, \relax
\mciteBstWouldAddEndPunctfalse
\mciteSetBstMidEndSepPunct{\mcitedefaultmidpunct}
{}{\mcitedefaultseppunct}\relax
\EndOfBibitem
\bibitem[Major(2007)]{Major2007}
Major,~F.~G. In \emph{The Quantum Beat: Principles and Applications of Atomic
  Clocks}; Major,~F.~G., Ed.; Springer: New York, NY, 2007; pp 417--443\relax
\mciteBstWouldAddEndPuncttrue
\mciteSetBstMidEndSepPunct{\mcitedefaultmidpunct}
{\mcitedefaultendpunct}{\mcitedefaultseppunct}\relax
\EndOfBibitem
\bibitem[Ludlow \latin{et~al.}(2015)Ludlow, Boyd, Ye, Peik, and
  Schmidt]{Ludlow2015}
Ludlow,~A.~D.; Boyd,~M.~M.; Ye,~J.; Peik,~E.; Schmidt,~P.~O. Optical atomic
  clocks. \emph{Rev. Mod. Phys.} \textbf{2015}, \emph{87}, 637--701\relax
\mciteBstWouldAddEndPuncttrue
\mciteSetBstMidEndSepPunct{\mcitedefaultmidpunct}
{\mcitedefaultendpunct}{\mcitedefaultseppunct}\relax
\EndOfBibitem
\bibitem[Fischer \latin{et~al.}(2004)Fischer, Kolachevsky, Zimmermann,
  Holzwarth, Udem, Hänsch, Abgrall, Grünert, Maksimovic, Bize, Marion,
  Santos, Lemonde, Santarelli, Laurent, Clairon, Salomon, Haas, Jentschura, and
  Keitel]{Fischer2004}
Fischer,~M.; Kolachevsky,~N.; Zimmermann,~M.; Holzwarth,~R.; Udem,~T.;
  Hänsch,~T.~W.; Abgrall,~M.; Grünert,~J.; Maksimovic,~I.; Bize,~S.
  \latin{et~al.}  New {Limits} on the {Drift} of {Fundamental} {Constants} from
  {Laboratory} {Measurements}. \emph{Phys. Rev. Lett.} \textbf{2004},
  \emph{92}, 230802\relax
\mciteBstWouldAddEndPuncttrue
\mciteSetBstMidEndSepPunct{\mcitedefaultmidpunct}
{\mcitedefaultendpunct}{\mcitedefaultseppunct}\relax
\EndOfBibitem
\bibitem[Blatt \latin{et~al.}(2008)Blatt, Ludlow, Campbell, Thomsen,
  Zelevinsky, Boyd, Ye, Baillard, Fouché, Le~Targat, Brusch, Lemonde,
  Takamoto, Hong, Katori, and Flambaum]{Blatt2008}
Blatt,~S.; Ludlow,~A.~D.; Campbell,~G.~K.; Thomsen,~J.~W.; Zelevinsky,~T.;
  Boyd,~M.~M.; Ye,~J.; Baillard,~X.; Fouché,~M.; Le~Targat,~R. \latin{et~al.}
  New {Limits} on {Coupling} of {Fundamental} {Constants} to {Gravity} {Using}
  \${\textasciicircum}\{87\}{\textbackslash}mathrm\{{Sr}\}\$ {Optical}
  {Lattice} {Clocks}. \emph{Phys. Rev. Lett.} \textbf{2008}, \emph{100},
  140801\relax
\mciteBstWouldAddEndPuncttrue
\mciteSetBstMidEndSepPunct{\mcitedefaultmidpunct}
{\mcitedefaultendpunct}{\mcitedefaultseppunct}\relax
\EndOfBibitem
\bibitem[Rosenband \latin{et~al.}(2008)Rosenband, Hume, Schmidt, Chou, Brusch,
  Lorini, Oskay, Drullinger, Fortier, Stalnaker, Diddams, Swann, Newbury,
  Itano, Wineland, and Bergquist]{Rosenband2008}
Rosenband,~T.; Hume,~D.~B.; Schmidt,~P.~O.; Chou,~C.~W.; Brusch,~A.;
  Lorini,~L.; Oskay,~W.~H.; Drullinger,~R.~E.; Fortier,~T.~M.; Stalnaker,~J.~E.
  \latin{et~al.}  Frequency {Ratio} of {Al}+ and {Hg}+ {Single}-{Ion} {Optical}
  {Clocks}; {Metrology} at the 17th {Decimal} {Place}. \emph{Science}
  \textbf{2008}, \emph{319}, 1808--1812\relax
\mciteBstWouldAddEndPuncttrue
\mciteSetBstMidEndSepPunct{\mcitedefaultmidpunct}
{\mcitedefaultendpunct}{\mcitedefaultseppunct}\relax
\EndOfBibitem
\bibitem[Dzuba \latin{et~al.}(2021)Dzuba, Allehabi, Flambaum, Li, and
  Schiller]{Dzuba2021}
Dzuba,~V.~A.; Allehabi,~S.~O.; Flambaum,~V.~V.; Li,~J.; Schiller,~S. Time
  Keeping and Searching for New Physics Using Metastable States of {{Cu}},
  {{Ag}}, and {{Au}}. \emph{Phys. Rev. A} \textbf{2021}, \emph{103},
  022822\relax
\mciteBstWouldAddEndPuncttrue
\mciteSetBstMidEndSepPunct{\mcitedefaultmidpunct}
{\mcitedefaultendpunct}{\mcitedefaultseppunct}\relax
\EndOfBibitem
\bibitem[Safronova \latin{et~al.}(2012)Safronova, Kozlov, and
  Clark]{Safronova2012a}
Safronova,~M.~S.; Kozlov,~M.~G.; Clark,~C.~W. Blackbody radiation shifts in
  optical atomic clocks. \emph{IEEE Trans. Ultrason. Ferroelectr. Freq.
  Control} \textbf{2012}, \emph{59}, 439--447\relax
\mciteBstWouldAddEndPuncttrue
\mciteSetBstMidEndSepPunct{\mcitedefaultmidpunct}
{\mcitedefaultendpunct}{\mcitedefaultseppunct}\relax
\EndOfBibitem
\bibitem[Kozlov \latin{et~al.}(2001)Kozlov, Porsev, and Johnson]{Kozlov2001}
Kozlov,~M.~G.; Porsev,~S.~G.; Johnson,~W.~R. Parity Nonconservation in
  Thallium. \emph{Phys. Rev. A} \textbf{2001}, \emph{64}, 052107\relax
\mciteBstWouldAddEndPuncttrue
\mciteSetBstMidEndSepPunct{\mcitedefaultmidpunct}
{\mcitedefaultendpunct}{\mcitedefaultseppunct}\relax
\EndOfBibitem
\bibitem[Porsev and Derevianko(2006)Porsev, and Derevianko]{Porsev2006}
Porsev,~S.~G.; Derevianko,~A. High-Accuracy Calculations of Dipole, Quadrupole,
  and Octupole Electric Dynamic Polarizabilities and van Der {{Waals}}
  Coefficients {{C6}}, {{C8}}, and {{C10}} for Alkaline-Earth Dimers. \emph{J.
  Exp. Theor. Phys.} \textbf{2006}, \emph{102}, 195--205\relax
\mciteBstWouldAddEndPuncttrue
\mciteSetBstMidEndSepPunct{\mcitedefaultmidpunct}
{\mcitedefaultendpunct}{\mcitedefaultseppunct}\relax
\EndOfBibitem
\bibitem[Porsev and Derevianko(2006)Porsev, and Derevianko]{Porsev2006a}
Porsev,~S.~G.; Derevianko,~A. Multipolar Theory of Blackbody Radiation Shift of
  Atomic Energy Levels and Its Implications for Optical Lattice Clocks.
  \emph{Phys. Rev. A} \textbf{2006}, \emph{74}, 020502\relax
\mciteBstWouldAddEndPuncttrue
\mciteSetBstMidEndSepPunct{\mcitedefaultmidpunct}
{\mcitedefaultendpunct}{\mcitedefaultseppunct}\relax
\EndOfBibitem
\bibitem[Dzuba and Flambaum(2009)Dzuba, and Flambaum]{Dzuba2009}
Dzuba,~V.~A.; Flambaum,~V.~V. Calculation of the \$({{T}},{{P}})\$-Odd Electric
  Dipole Moment of Thallium and Cesium. \emph{Phys. Rev. A} \textbf{2009},
  \emph{80}, 062509\relax
\mciteBstWouldAddEndPuncttrue
\mciteSetBstMidEndSepPunct{\mcitedefaultmidpunct}
{\mcitedefaultendpunct}{\mcitedefaultseppunct}\relax
\EndOfBibitem
\bibitem[Mitroy \latin{et~al.}(2010)Mitroy, Safronova, and Clark]{Mitroy2010a}
Mitroy,~J.; Safronova,~M.~S.; Clark,~C.~W. Theory and Applications of Atomic
  and Ionic Polarizabilities. \emph{J. Phys. B: At. Mol. Opt. Phys.}
  \textbf{2010}, \emph{43}, 202001\relax
\mciteBstWouldAddEndPuncttrue
\mciteSetBstMidEndSepPunct{\mcitedefaultmidpunct}
{\mcitedefaultendpunct}{\mcitedefaultseppunct}\relax
\EndOfBibitem
\bibitem[Blundell \latin{et~al.}(1989)Blundell, Johnson, Liu, and
  Sapirstein]{Blundell1989}
Blundell,~S.~A.; Johnson,~W.~R.; Liu,~Z.~W.; Sapirstein,~J. Relativistic
  all-order calculations of energies and matrix elements for {Li} and
  \$\{{\textbackslash}mathrm\{{Be}\}\}{\textasciicircum}\{+\}\$. \emph{Physical
  Review A} \textbf{1989}, \emph{40}, 2233--2246, Publisher: American Physical
  Society\relax
\mciteBstWouldAddEndPuncttrue
\mciteSetBstMidEndSepPunct{\mcitedefaultmidpunct}
{\mcitedefaultendpunct}{\mcitedefaultseppunct}\relax
\EndOfBibitem
\bibitem[Derevianko \latin{et~al.}(1999)Derevianko, Johnson, Safronova, and
  Babb]{Derevianko1999}
Derevianko,~A.; Johnson,~W.~R.; Safronova,~M.~S.; Babb,~J.~F. High-{{Precision
  Calculations}} of {{Dispersion Coefficients}}, {{Static Dipole
  Polarizabilities}}, and {{Atom-Wall Interaction Constants}} for
  {{Alkali-Metal Atoms}}. \emph{Phys. Rev. Lett.} \textbf{1999}, \emph{82},
  3589--3592\relax
\mciteBstWouldAddEndPuncttrue
\mciteSetBstMidEndSepPunct{\mcitedefaultmidpunct}
{\mcitedefaultendpunct}{\mcitedefaultseppunct}\relax
\EndOfBibitem
\bibitem[Derevianko \latin{et~al.}(2010)Derevianko, Porsev, and
  Babb]{Derevianko2010}
Derevianko,~A.; Porsev,~S.~G.; Babb,~J.~F. Electric dipole polarizabilities at
  imaginary frequencies for hydrogen, the alkali–metal, alkaline–earth, and
  noble gas atoms. \emph{Atomic Data and Nuclear Data Tables} \textbf{2010},
  \emph{96}, 323--331\relax
\mciteBstWouldAddEndPuncttrue
\mciteSetBstMidEndSepPunct{\mcitedefaultmidpunct}
{\mcitedefaultendpunct}{\mcitedefaultseppunct}\relax
\EndOfBibitem
\bibitem[Safronova \latin{et~al.}(2007)Safronova, Johnson, and
  Safronova]{Safronova2007}
Safronova,~U.~I.; Johnson,~W.~R.; Safronova,~M.~S. Excitation Energies,
  Polarizabilities, Multipole Transition Rates, and Lifetimes of Ions along the
  Francium Isoelectronic Sequence. \emph{Phys. Rev. A} \textbf{2007},
  \emph{76}, 042504\relax
\mciteBstWouldAddEndPuncttrue
\mciteSetBstMidEndSepPunct{\mcitedefaultmidpunct}
{\mcitedefaultendpunct}{\mcitedefaultseppunct}\relax
\EndOfBibitem
\bibitem[Johnson \latin{et~al.}(2008)Johnson, Safronova, Derevianko, and
  Safronova]{Johnson2008}
Johnson,~W.~R.; Safronova,~U.~I.; Derevianko,~A.; Safronova,~M.~S. Relativistic
  Many-Body Calculation of Energies, Lifetimes, Hyperfine Constants, and
  Polarizabilities in
  \${\textasciicircum}\{7\}{\textbackslash}text\{\vphantom\}{{L}}\vphantom\{\}{\textbackslash}text\{i\}\$.
  \emph{Phys. Rev. A} \textbf{2008}, \emph{77}, 022510\relax
\mciteBstWouldAddEndPuncttrue
\mciteSetBstMidEndSepPunct{\mcitedefaultmidpunct}
{\mcitedefaultendpunct}{\mcitedefaultseppunct}\relax
\EndOfBibitem
\bibitem[Badhan \latin{et~al.}(2022)Badhan, Kaur, Arora, and Sahoo]{Badhan2022}
Badhan,~V.; Kaur,~S.; Arora,~B.; Sahoo,~B.~K. Assessing Slowdown Times Due to
  Blackbody Friction Forces for High-Precision Experiments. \emph{Eur. Phys. J.
  D} \textbf{2022}, \emph{76}, 252\relax
\mciteBstWouldAddEndPuncttrue
\mciteSetBstMidEndSepPunct{\mcitedefaultmidpunct}
{\mcitedefaultendpunct}{\mcitedefaultseppunct}\relax
\EndOfBibitem
\bibitem[Cheng(2024)]{Cheng2024a}
Cheng,~Y. Relativistic and electron-correlation effects in static dipole
  polarizabilities for main-group elements. \emph{Phys. Rev. A} \textbf{2024},
  \emph{110}, 042805\relax
\mciteBstWouldAddEndPuncttrue
\mciteSetBstMidEndSepPunct{\mcitedefaultmidpunct}
{\mcitedefaultendpunct}{\mcitedefaultseppunct}\relax
\EndOfBibitem
\bibitem[Dyall(2002)]{Dyall2002}
Dyall,~K.~G. Relativistic and Nonrelativistic Finite Nucleus Optimized
  Triple-Zeta Basis Sets for the 4p, 5p and 6p Elements. \emph{Theor. Chem.
  Acc.} \textbf{2002}, \emph{108}, 335--340\relax
\mciteBstWouldAddEndPuncttrue
\mciteSetBstMidEndSepPunct{\mcitedefaultmidpunct}
{\mcitedefaultendpunct}{\mcitedefaultseppunct}\relax
\EndOfBibitem
\bibitem[Dyall(2004)]{Dyall2004}
Dyall,~K.~G. Relativistic Double-Zeta, Triple-Zeta, and Quadruple-Zeta Basis
  Sets for the 5d Elements {{Hf}}\textendash{{Hg}}. \emph{Theor. Chem. Acc.}
  \textbf{2004}, \emph{112}, 403--409\relax
\mciteBstWouldAddEndPuncttrue
\mciteSetBstMidEndSepPunct{\mcitedefaultmidpunct}
{\mcitedefaultendpunct}{\mcitedefaultseppunct}\relax
\EndOfBibitem
\bibitem[Dyall(2006)]{Dyall2006}
Dyall,~K.~G. Relativistic {{Quadruple-Zeta}} and {{Revised Triple-Zeta}} and
  {{Double-Zeta Basis Sets}} for the 4p, 5p, and 6p {{Elements}}. \emph{Theor.
  Chem. Acc.} \textbf{2006}, \emph{115}, 441--447\relax
\mciteBstWouldAddEndPuncttrue
\mciteSetBstMidEndSepPunct{\mcitedefaultmidpunct}
{\mcitedefaultendpunct}{\mcitedefaultseppunct}\relax
\EndOfBibitem
\bibitem[Dyall(2007)]{Dyall2007}
Dyall,~K.~G. Relativistic Double-Zeta, Triple-Zeta, and Quadruple-Zeta Basis
  Sets for the 4d Elements {{Y}}\textendash{{Cd}}. \emph{Theor. Chem. Acc.}
  \textbf{2007}, \emph{117}, 483--489\relax
\mciteBstWouldAddEndPuncttrue
\mciteSetBstMidEndSepPunct{\mcitedefaultmidpunct}
{\mcitedefaultendpunct}{\mcitedefaultseppunct}\relax
\EndOfBibitem
\bibitem[Dyall(2009)]{Dyall2009}
Dyall,~K.~G. Relativistic {{Double-Zeta}}, {{Triple-Zeta}}, and
  {{Quadruple-Zeta Basis Sets}} for the 4s, 5s, 6s, and 7s {{Elements}}.
  \emph{J. Phys. Chem. A} \textbf{2009}, \emph{113}, 12638--12644\relax
\mciteBstWouldAddEndPuncttrue
\mciteSetBstMidEndSepPunct{\mcitedefaultmidpunct}
{\mcitedefaultendpunct}{\mcitedefaultseppunct}\relax
\EndOfBibitem
\bibitem[Dyall and Gomes(2010)Dyall, and Gomes]{Dyall2010}
Dyall,~K.~G.; Gomes,~A. S.~P. Revised Relativistic Basis Sets for the 5d
  Elements {{Hf}}\textendash{{Hg}}. \emph{Theor. Chem. Acc.} \textbf{2010},
  \emph{125}, 97--100\relax
\mciteBstWouldAddEndPuncttrue
\mciteSetBstMidEndSepPunct{\mcitedefaultmidpunct}
{\mcitedefaultendpunct}{\mcitedefaultseppunct}\relax
\EndOfBibitem
\bibitem[Roos \latin{et~al.}(2004)Roos, Lindh, Malmqvist, Veryazov, and
  Widmark]{Roos2004}
Roos,~B.~O.; Lindh,~R.; Malmqvist,~P.-{\AA}.; Veryazov,~V.; Widmark,~P.-O. Main
  {{Group Atoms}} and {{Dimers Studied}} with a {{New Relativistic ANO Basis
  Set}}. \emph{J. Phys. Chem. A} \textbf{2004}, \emph{108}, 2851--2858\relax
\mciteBstWouldAddEndPuncttrue
\mciteSetBstMidEndSepPunct{\mcitedefaultmidpunct}
{\mcitedefaultendpunct}{\mcitedefaultseppunct}\relax
\EndOfBibitem
\bibitem[Roos \latin{et~al.}(2005)Roos, Lindh, Malmqvist, Veryazov, and
  Widmark]{Roos2005}
Roos,~B.~O.; Lindh,~R.; Malmqvist,~P.-{\AA}.~{\AA}.; Veryazov,~V.;
  Widmark,~P.-O.~O. New {{Relativistic ANO Basis Sets}} for {{Transition Metal
  Atoms}}. \emph{J. Phys. Chem. A} \textbf{2005}, \emph{109}, 6575--6579\relax
\mciteBstWouldAddEndPuncttrue
\mciteSetBstMidEndSepPunct{\mcitedefaultmidpunct}
{\mcitedefaultendpunct}{\mcitedefaultseppunct}\relax
\EndOfBibitem
\bibitem[Neogr{\'a}dy \latin{et~al.}(1997)Neogr{\'a}dy, Kell{\"o}, Urban, and
  Sadlej]{Neogrády1997}
Neogr{\'a}dy,~P.; Kell{\"o},~V.; Urban,~M.; Sadlej,~A.~J. Ionization Potentials
  and Electron Affinities of {{Cu}}, {{Ag}}, and {{Au}}: {{Electron}}
  Correlation and Relativistic Effects. \emph{Int. J. Quantum Chem.}
  \textbf{1997}, \emph{63}, 557--565\relax
\mciteBstWouldAddEndPuncttrue
\mciteSetBstMidEndSepPunct{\mcitedefaultmidpunct}
{\mcitedefaultendpunct}{\mcitedefaultseppunct}\relax
\EndOfBibitem
\bibitem[Maroulis(2006)]{Maroulis2006}
Maroulis,~G. \emph{Atoms, {{Molecules And Clusters In Electric Fields}}:
  {{Theoretical Approaches To The Calculation Of Electric Polarizability}}};
  World Scientific, 2006\relax
\mciteBstWouldAddEndPuncttrue
\mciteSetBstMidEndSepPunct{\mcitedefaultmidpunct}
{\mcitedefaultendpunct}{\mcitedefaultseppunct}\relax
\EndOfBibitem
\bibitem[Mohr(2009)]{Mohr2009}
Mohr,~F. \emph{Gold {{Chemistry}}: {{Applications}} and {{Future Directions}}
  in the {{Life Sciences}}}; John Wiley \& Sons, 2009\relax
\mciteBstWouldAddEndPuncttrue
\mciteSetBstMidEndSepPunct{\mcitedefaultmidpunct}
{\mcitedefaultendpunct}{\mcitedefaultseppunct}\relax
\EndOfBibitem
\bibitem[Gobre(2016)]{gobre2016efficient}
Gobre,~V.~V. \emph{Efficient {{Modelling}} of {{Linear Electronic
  Polarization}} in {{Materials Using Atomic Response Functions}}}; Technische
  Universitaet Berlin (Germany), 2016\relax
\mciteBstWouldAddEndPuncttrue
\mciteSetBstMidEndSepPunct{\mcitedefaultmidpunct}
{\mcitedefaultendpunct}{\mcitedefaultseppunct}\relax
\EndOfBibitem
\bibitem[Tomza(2021)]{Tomza2021}
Tomza,~M. Interaction Potentials, Electric Moments, Polarizabilities, and
  Chemical Reactions of {{YbCu}}, {{YbAg}}, and {{YbAu}} Molecules. \emph{New
  J. Phys.} \textbf{2021}, \emph{23}, 085003\relax
\mciteBstWouldAddEndPuncttrue
\mciteSetBstMidEndSepPunct{\mcitedefaultmidpunct}
{\mcitedefaultendpunct}{\mcitedefaultseppunct}\relax
\EndOfBibitem
\bibitem[{\'S}mia{\l}kowski and Tomza(2021){\'S}mia{\l}kowski, and
  Tomza]{Smialkowski2021}
{\'S}mia{\l}kowski,~M.; Tomza,~M. Highly Polar Molecules Consisting of a Copper
  or Silver Atom Interacting with an Alkali-Metal or Alkaline-Earth-Metal Atom.
  \emph{Phys. Rev. A} \textbf{2021}, \emph{103}, 022802\relax
\mciteBstWouldAddEndPuncttrue
\mciteSetBstMidEndSepPunct{\mcitedefaultmidpunct}
{\mcitedefaultendpunct}{\mcitedefaultseppunct}\relax
\EndOfBibitem
\bibitem[Wesendrup and Schwerdtfeger(2000)Wesendrup, and
  Schwerdtfeger]{Wesendrup2000}
Wesendrup,~R.; Schwerdtfeger,~P. Extremely {{Strong}} S2 -- S2 {{Closed-Shell
  Interactions}}. \emph{Angew. Chem. Int. Ed.} \textbf{2000}, \emph{39},
  907--910\relax
\mciteBstWouldAddEndPuncttrue
\mciteSetBstMidEndSepPunct{\mcitedefaultmidpunct}
{\mcitedefaultendpunct}{\mcitedefaultseppunct}\relax
\EndOfBibitem
\bibitem[Gould and Bu{\v c}ko(2016)Gould, and Bu{\v c}ko]{Gould2016a}
Gould,~T.; Bu{\v c}ko,~T. C6 {{Coefficients}} and {{Dipole Polarizabilities}}
  for {{All Atoms}} and {{Many Ions}} in {{Rows}} 1--6 of the {{Periodic
  Table}}. \emph{J. Chem. Theory Comput.} \textbf{2016}, \emph{12},
  3603--3613\relax
\mciteBstWouldAddEndPuncttrue
\mciteSetBstMidEndSepPunct{\mcitedefaultmidpunct}
{\mcitedefaultendpunct}{\mcitedefaultseppunct}\relax
\EndOfBibitem
\bibitem[Yu \latin{et~al.}(2015)Yu, Suo, Feng, Fan, and Liu]{Yu2015}
Yu,~Y.-m.; Suo,~B.-b.; Feng,~H.-h.; Fan,~H.; Liu,~W.-M. Finite-field
  calculation of static polarizabilities and hyperpolarizabilities of
  ${\text{In}}^{+}$ and Sr. \emph{Phys. Rev. A} \textbf{2015}, \emph{92},
  052515\relax
\mciteBstWouldAddEndPuncttrue
\mciteSetBstMidEndSepPunct{\mcitedefaultmidpunct}
{\mcitedefaultendpunct}{\mcitedefaultseppunct}\relax
\EndOfBibitem
\bibitem[Fleig(2012)]{Fleig2012}
Fleig,~T. Invited Review: {{Relativistic}} Wave-Function Based Electron
  Correlation Methods. \emph{Chem. Phys.} \textbf{2012}, \emph{395},
  2--15\relax
\mciteBstWouldAddEndPuncttrue
\mciteSetBstMidEndSepPunct{\mcitedefaultmidpunct}
{\mcitedefaultendpunct}{\mcitedefaultseppunct}\relax
\EndOfBibitem
\bibitem[Saue(2011)]{Saue2011}
Saue,~T. Relativistic {{Hamiltonians}} for {{Chemistry}}: {{A Primer}}.
  \emph{ChemPhysChem} \textbf{2011}, \emph{12}, 3077--3094\relax
\mciteBstWouldAddEndPuncttrue
\mciteSetBstMidEndSepPunct{\mcitedefaultmidpunct}
{\mcitedefaultendpunct}{\mcitedefaultseppunct}\relax
\EndOfBibitem
\bibitem[Douglas and Kroll(1974)Douglas, and Kroll]{Douglas1974}
Douglas,~M.; Kroll,~N.~M. Quantum Electrodynamical Corrections to the Fine
  Structure of Helium. \emph{Ann. Phys.} \textbf{1974}, \emph{82},
  89--155\relax
\mciteBstWouldAddEndPuncttrue
\mciteSetBstMidEndSepPunct{\mcitedefaultmidpunct}
{\mcitedefaultendpunct}{\mcitedefaultseppunct}\relax
\EndOfBibitem
\bibitem[Hess(1985)]{Hess1985}
Hess,~B.~A. Applicability of the No-Pair Equation with Free-Particle Projection
  Operators to Atomic and Molecular Structure Calculations. \emph{Phys. Rev. A}
  \textbf{1985}, \emph{32}, 756--763\relax
\mciteBstWouldAddEndPuncttrue
\mciteSetBstMidEndSepPunct{\mcitedefaultmidpunct}
{\mcitedefaultendpunct}{\mcitedefaultseppunct}\relax
\EndOfBibitem
\bibitem[Hess(1986)]{Hess1986}
Hess,~B.~A. Relativistic Electronic-Structure Calculations Employing a
  Two-Component No-Pair Formalism with External-Field Projection Operators.
  \emph{Phys. Rev. A} \textbf{1986}, \emph{33}, 3742--3748\relax
\mciteBstWouldAddEndPuncttrue
\mciteSetBstMidEndSepPunct{\mcitedefaultmidpunct}
{\mcitedefaultendpunct}{\mcitedefaultseppunct}\relax
\EndOfBibitem
\bibitem[Chang \latin{et~al.}(1986)Chang, Pelissier, and Durand]{Chang1986}
Chang,~C.; Pelissier,~M.; Durand,~P. Regular {{Two-Component Pauli-Like
  Effective Hamiltonians}} in {{Dirac Theory}}. \emph{Phys. Scr.}
  \textbf{1986}, \emph{34}, 394\relax
\mciteBstWouldAddEndPuncttrue
\mciteSetBstMidEndSepPunct{\mcitedefaultmidpunct}
{\mcitedefaultendpunct}{\mcitedefaultseppunct}\relax
\EndOfBibitem
\bibitem[{van Lenthe} \latin{et~al.}(1994){van Lenthe}, Baerends, and
  Snijders]{vanLenthe1994}
{van Lenthe},~E.; Baerends,~E.~J.; Snijders,~J.~G. Relativistic Total Energy
  Using Regular Approximations. \emph{J. Chem. Phys.} \textbf{1994},
  \emph{101}, 9783--9792\relax
\mciteBstWouldAddEndPuncttrue
\mciteSetBstMidEndSepPunct{\mcitedefaultmidpunct}
{\mcitedefaultendpunct}{\mcitedefaultseppunct}\relax
\EndOfBibitem
\bibitem[{van Lenthe} \latin{et~al.}(1996){van Lenthe}, Snijders, and
  Baerends]{vanLenthe1996}
{van Lenthe},~E.; Snijders,~J.~G.; Baerends,~E.~J. The Zero-order Regular
  Approximation for Relativistic Effects: {{The}} Effect of Spin\textendash
  Orbit Coupling in Closed Shell Molecules. \emph{J. Chem. Phys.}
  \textbf{1996}, \emph{105}, 6505--6516\relax
\mciteBstWouldAddEndPuncttrue
\mciteSetBstMidEndSepPunct{\mcitedefaultmidpunct}
{\mcitedefaultendpunct}{\mcitedefaultseppunct}\relax
\EndOfBibitem
\bibitem[Ilia{\v s} and Saue(2007)Ilia{\v s}, and Saue]{Ilias2007}
Ilia{\v s},~M.; Saue,~T. An Infinite-Order Two-Component Relativistic
  {{Hamiltonian}} by a Simple One-Step Transformation. \emph{J. Chem. Phys.}
  \textbf{2007}, \emph{126}, 064102\relax
\mciteBstWouldAddEndPuncttrue
\mciteSetBstMidEndSepPunct{\mcitedefaultmidpunct}
{\mcitedefaultendpunct}{\mcitedefaultseppunct}\relax
\EndOfBibitem
\bibitem[Dyall(1997)]{Dyall1997}
Dyall,~K.~G. Interfacing relativistic and nonrelativistic methods. {I}.
  {Normalized} elimination of the small component in the modified {Dirac}
  equation. \emph{The Journal of Chemical Physics} \textbf{1997}, \emph{106},
  9618--9626\relax
\mciteBstWouldAddEndPuncttrue
\mciteSetBstMidEndSepPunct{\mcitedefaultmidpunct}
{\mcitedefaultendpunct}{\mcitedefaultseppunct}\relax
\EndOfBibitem
\bibitem[Dyall(2002)]{Dyall2002a}
Dyall,~K.~G. A systematic sequence of relativistic approximations.
  \emph{Journal of Computational Chemistry} \textbf{2002}, \emph{23}, 786--793,
  \_eprint: https://onlinelibrary.wiley.com/doi/pdf/10.1002/jcc.10048\relax
\mciteBstWouldAddEndPuncttrue
\mciteSetBstMidEndSepPunct{\mcitedefaultmidpunct}
{\mcitedefaultendpunct}{\mcitedefaultseppunct}\relax
\EndOfBibitem
\bibitem[Kutzelnigg and Liu(2005)Kutzelnigg, and Liu]{Kutzelnigg2005}
Kutzelnigg,~W.; Liu,~W. Quasirelativistic theory equivalent to fully
  relativistic theory. \emph{The Journal of Chemical Physics} \textbf{2005},
  \emph{123}, 241102\relax
\mciteBstWouldAddEndPuncttrue
\mciteSetBstMidEndSepPunct{\mcitedefaultmidpunct}
{\mcitedefaultendpunct}{\mcitedefaultseppunct}\relax
\EndOfBibitem
\bibitem[Filatov and Dyall(2007)Filatov, and Dyall]{Filatov2007}
Filatov,~M.; Dyall,~K.~G. On convergence of the normalized elimination of the
  small component ({NESC}) method. \emph{Theoretical Chemistry Accounts}
  \textbf{2007}, \emph{117}, 333--338\relax
\mciteBstWouldAddEndPuncttrue
\mciteSetBstMidEndSepPunct{\mcitedefaultmidpunct}
{\mcitedefaultendpunct}{\mcitedefaultseppunct}\relax
\EndOfBibitem
\bibitem[Peng \latin{et~al.}(2007)Peng, Liu, Xiao, and Cheng]{Peng2007}
Peng,~D.; Liu,~W.; Xiao,~Y.; Cheng,~L. Making four- and two-component
  relativistic density functional methods fully equivalent based on the idea of
  “from atoms to molecule”. \emph{The Journal of Chemical Physics}
  \textbf{2007}, \emph{127}, 104106\relax
\mciteBstWouldAddEndPuncttrue
\mciteSetBstMidEndSepPunct{\mcitedefaultmidpunct}
{\mcitedefaultendpunct}{\mcitedefaultseppunct}\relax
\EndOfBibitem
\bibitem[Liu and Peng(2009)Liu, and Peng]{Liu2009}
Liu,~W.; Peng,~D. Exact two-component {Hamiltonians} revisited. \emph{The
  Journal of Chemical Physics} \textbf{2009}, \emph{131}, 031104\relax
\mciteBstWouldAddEndPuncttrue
\mciteSetBstMidEndSepPunct{\mcitedefaultmidpunct}
{\mcitedefaultendpunct}{\mcitedefaultseppunct}\relax
\EndOfBibitem
\bibitem[Peng and Reiher(2012)Peng, and Reiher]{Peng2012}
Peng,~D.; Reiher,~M. Exact decoupling of the relativistic {Fock} operator.
  \emph{Theoretical Chemistry Accounts} \textbf{2012}, \emph{131}, 1081\relax
\mciteBstWouldAddEndPuncttrue
\mciteSetBstMidEndSepPunct{\mcitedefaultmidpunct}
{\mcitedefaultendpunct}{\mcitedefaultseppunct}\relax
\EndOfBibitem
\bibitem[Liu(2014)]{Liu2014}
Liu,~W. Advances in relativistic molecular quantum mechanics. \emph{Physics
  Reports} \textbf{2014}, \emph{537}, 59--89\relax
\mciteBstWouldAddEndPuncttrue
\mciteSetBstMidEndSepPunct{\mcitedefaultmidpunct}
{\mcitedefaultendpunct}{\mcitedefaultseppunct}\relax
\EndOfBibitem
\bibitem[Liu(2016)]{Liu2016a}
Liu,~W. Big picture of relativistic molecular quantum mechanics. \emph{National
  Science Review} \textbf{2016}, \emph{3}, 204--221\relax
\mciteBstWouldAddEndPuncttrue
\mciteSetBstMidEndSepPunct{\mcitedefaultmidpunct}
{\mcitedefaultendpunct}{\mcitedefaultseppunct}\relax
\EndOfBibitem
\bibitem[Liu(2020)]{Liu2020}
Liu,~W. Essentials of relativistic quantum chemistry. \emph{The Journal of
  Chemical Physics} \textbf{2020}, \emph{152}, 180901\relax
\mciteBstWouldAddEndPuncttrue
\mciteSetBstMidEndSepPunct{\mcitedefaultmidpunct}
{\mcitedefaultendpunct}{\mcitedefaultseppunct}\relax
\EndOfBibitem
\bibitem[Dyall(2001)]{Dyall2001}
Dyall,~K.~G. Interfacing relativistic and nonrelativistic methods. {IV}. {One}-
  and two-electron scalar approximations. \emph{The Journal of Chemical
  Physics} \textbf{2001}, \emph{115}, 9136--9143\relax
\mciteBstWouldAddEndPuncttrue
\mciteSetBstMidEndSepPunct{\mcitedefaultmidpunct}
{\mcitedefaultendpunct}{\mcitedefaultseppunct}\relax
\EndOfBibitem
\bibitem[Saue \latin{et~al.}(2020)Saue, Bast, Gomes, Jensen, Visscher, Aucar,
  Di~Remigio, Dyall, Eliav, Fasshauer, Fleig, Halbert, Hedeg{\aa}rd,
  {Helmich-Paris}, Ilia{\v s}, Jacob, Knecht, Laerdahl, Vidal, Nayak,
  Olejniczak, Olsen, Pernpointner, Senjean, Shee, Sunaga, and {van
  Stralen}]{Saue2020}
Saue,~T.; Bast,~R.; Gomes,~A. S.~P.; Jensen,~H. J.~A.; Visscher,~L.;
  Aucar,~I.~A.; Di~Remigio,~R.; Dyall,~K.~G.; Eliav,~E.; Fasshauer,~E.
  \latin{et~al.}  The {{DIRAC}} Code for Relativistic Molecular Calculations.
  \emph{J. Chem. Phys.} \textbf{2020}, \emph{152}, 204104\relax
\mciteBstWouldAddEndPuncttrue
\mciteSetBstMidEndSepPunct{\mcitedefaultmidpunct}
{\mcitedefaultendpunct}{\mcitedefaultseppunct}\relax
\EndOfBibitem
\bibitem[{van Stralen} \latin{et~al.}(2005){van Stralen}, Visscher, Larsen, and
  Jensen]{vanStralen2005a}
{van Stralen},~J. N.~P.; Visscher,~L.; Larsen,~C.~V.; Jensen,~H. J.~A.
  First-Order {{MP2}} Molecular Properties in a Relativistic Framework.
  \emph{Chem. Phys.} \textbf{2005}, \emph{311}, 81--95\relax
\mciteBstWouldAddEndPuncttrue
\mciteSetBstMidEndSepPunct{\mcitedefaultmidpunct}
{\mcitedefaultendpunct}{\mcitedefaultseppunct}\relax
\EndOfBibitem
\bibitem[Visscher \latin{et~al.}(1996)Visscher, Lee, and Dyall]{Visscher1996}
Visscher,~L.; Lee,~T.~J.; Dyall,~K.~G. Formulation and Implementation of a
  Relativistic Unrestricted Coupled-cluster Method Including Noniterative
  Connected Triples. \emph{J. Chem. Phys.} \textbf{1996}, \emph{105},
  8769--8776\relax
\mciteBstWouldAddEndPuncttrue
\mciteSetBstMidEndSepPunct{\mcitedefaultmidpunct}
{\mcitedefaultendpunct}{\mcitedefaultseppunct}\relax
\EndOfBibitem
\bibitem[Das and Thakkar(1998)Das, and Thakkar]{Das1998}
Das,~A.~K.; Thakkar,~A.~J. Static Response Properties of Second-Period Atoms:
  Coupled Cluster Calculations. \emph{J. Phys. B At. Mol. Opt. Phys.}
  \textbf{1998}, \emph{31}, 2215\relax
\mciteBstWouldAddEndPuncttrue
\mciteSetBstMidEndSepPunct{\mcitedefaultmidpunct}
{\mcitedefaultendpunct}{\mcitedefaultseppunct}\relax
\EndOfBibitem
\bibitem[Williams(2016)]{Williams2016}
Williams,~J.~H. \emph{Quantifying {Measurement}: {The} tyranny of numbers};
  Morgan \& Claypool Publishers, 2016\relax
\mciteBstWouldAddEndPuncttrue
\mciteSetBstMidEndSepPunct{\mcitedefaultmidpunct}
{\mcitedefaultendpunct}{\mcitedefaultseppunct}\relax
\EndOfBibitem
\bibitem[pyd()]{pydirac}
pydirac 2024.7.8. \url{https://pypi.org/project/pydirac/}, Accessed:
  2024-07-08\relax
\mciteBstWouldAddEndPuncttrue
\mciteSetBstMidEndSepPunct{\mcitedefaultmidpunct}
{\mcitedefaultendpunct}{\mcitedefaultseppunct}\relax
\EndOfBibitem
\bibitem[Kállay \latin{et~al.}(2011)Kállay, Nataraj, Sahoo, Das, and
  Visscher]{Kallay2011}
Kállay,~M.; Nataraj,~H.~S.; Sahoo,~B.~K.; Das,~B.~P.; Visscher,~L.
  Relativistic general-order coupled-cluster method for high-precision
  calculations: {Application} to the {Al}\$\{\}{\textasciicircum}\{+\}\$ atomic
  clock. \emph{Phys. Rev. A} \textbf{2011}, \emph{83}, 030503\relax
\mciteBstWouldAddEndPuncttrue
\mciteSetBstMidEndSepPunct{\mcitedefaultmidpunct}
{\mcitedefaultendpunct}{\mcitedefaultseppunct}\relax
\EndOfBibitem
\bibitem[Irikura(2021)]{Irikura2021}
Irikura,~K.~K. Polarizability of atomic {Pt}, {Pt}+, and {Pt}-. \emph{J. Chem.
  Phys.} \textbf{2021}, \emph{154}, 174302\relax
\mciteBstWouldAddEndPuncttrue
\mciteSetBstMidEndSepPunct{\mcitedefaultmidpunct}
{\mcitedefaultendpunct}{\mcitedefaultseppunct}\relax
\EndOfBibitem
\bibitem[DIR()]{DIRAC18}
DIRAC18. \url{https://doi.org/10.5281/zenodo.2253986}, Accessed on Jun 7,
  2021\relax
\mciteBstWouldAddEndPuncttrue
\mciteSetBstMidEndSepPunct{\mcitedefaultmidpunct}
{\mcitedefaultendpunct}{\mcitedefaultseppunct}\relax
\EndOfBibitem
\bibitem[Sch()]{Schwerdtfeger2023}
2023 {Table} of static dipole polarizabilities of the neutral elements in the
  periodic table. \url{https://ctcp.massey.ac.nz/2023Tablepol.pdf}, Accessed on
  Jun 14, 2024\relax
\mciteBstWouldAddEndPuncttrue
\mciteSetBstMidEndSepPunct{\mcitedefaultmidpunct}
{\mcitedefaultendpunct}{\mcitedefaultseppunct}\relax
\EndOfBibitem
\bibitem[Schwerdtfeger and Bowmaker(1994)Schwerdtfeger, and
  Bowmaker]{Schwerdtfeger1994}
Schwerdtfeger,~P.; Bowmaker,~G.~A. Relativistic Effects in Gold Chemistry.
  {{V}}. {{Group}} 11 Dipole Polarizabilities and Weak Bonding in Monocarbonyl
  Compounds. \emph{J. Chem. Phys.} \textbf{1994}, \emph{100}, 4487--4497\relax
\mciteBstWouldAddEndPuncttrue
\mciteSetBstMidEndSepPunct{\mcitedefaultmidpunct}
{\mcitedefaultendpunct}{\mcitedefaultseppunct}\relax
\EndOfBibitem
\bibitem[{Pou-Am{\'e}rigo} \latin{et~al.}(1995){Pou-Am{\'e}rigo}, Merch{\'a}n,
  {Nebot-Gil}, Widmark, and Roos]{Pou-Amérigo1995}
{Pou-Am{\'e}rigo},~R.; Merch{\'a}n,~M.; {Nebot-Gil},~I.; Widmark,~P.-O.;
  Roos,~B.~O. Density Matrix Averaged Atomic Natural Orbital ({{ANO}}) Basis
  Sets for Correlated Molecular Wave Functions. \emph{Theoret. Chim. Acta}
  \textbf{1995}, \emph{92}, 149--181\relax
\mciteBstWouldAddEndPuncttrue
\mciteSetBstMidEndSepPunct{\mcitedefaultmidpunct}
{\mcitedefaultendpunct}{\mcitedefaultseppunct}\relax
\EndOfBibitem
\bibitem[Lide(2004)]{Lide2004}
Lide,~D.~R. \emph{{{CRC}} Handbook of Chemistry and Physics}; CRC press,
  2004\relax
\mciteBstWouldAddEndPuncttrue
\mciteSetBstMidEndSepPunct{\mcitedefaultmidpunct}
{\mcitedefaultendpunct}{\mcitedefaultseppunct}\relax
\EndOfBibitem
\bibitem[Doolen and Liberman(1987)Doolen, and Liberman]{Doolen1987}
Doolen,~G.; Liberman,~D.~A. Calculations of Photoabsorption by Atoms Using a
  Linear Response Method. \emph{Phys. Scr.} \textbf{1987}, \emph{36}, 77\relax
\mciteBstWouldAddEndPuncttrue
\mciteSetBstMidEndSepPunct{\mcitedefaultmidpunct}
{\mcitedefaultendpunct}{\mcitedefaultseppunct}\relax
\EndOfBibitem
\bibitem[Chu and Dalgarno(2004)Chu, and Dalgarno]{Chu2004}
Chu,~X.; Dalgarno,~A. Linear Response Time-Dependent Density Functional Theory
  for van Der {{Waals}} Coefficients. \emph{J. Chem. Phys.} \textbf{2004},
  \emph{121}, 4083--4088\relax
\mciteBstWouldAddEndPuncttrue
\mciteSetBstMidEndSepPunct{\mcitedefaultmidpunct}
{\mcitedefaultendpunct}{\mcitedefaultseppunct}\relax
\EndOfBibitem
\bibitem[K{\l}os(2005)]{Kłos2005a}
K{\l}os,~J. Anisotropic Dipole Polarizability of Transition Metal Atoms:
  {{Sc}}({{D2}}), {{Ti}}({{F3}},{{P3}}), {{V}}({{F4}},{{P4}},{{D6}}),
  {{Ni}}({{F3}}) and Ions: {{Sc2}}+({{D2}}), {{Ti2}}+({{F3}},{{P3}}). \emph{J.
  Chem. Phys.} \textbf{2005}, \emph{123}, 024308\relax
\mciteBstWouldAddEndPuncttrue
\mciteSetBstMidEndSepPunct{\mcitedefaultmidpunct}
{\mcitedefaultendpunct}{\mcitedefaultseppunct}\relax
\EndOfBibitem
\bibitem[Zhang \latin{et~al.}(2008)Zhang, Mitroy, Sadeghpour, and
  Bromley]{Zhang2008a}
Zhang,~J.~Y.; Mitroy,~J.; Sadeghpour,~H.~R.; Bromley,~M. W.~J. Long-Range
  Interactions of Copper and Silver Atoms with Hydrogen, Helium, and Rare-Gas
  Atoms. \emph{Phys. Rev. A} \textbf{2008}, \emph{78}, 062710\relax
\mciteBstWouldAddEndPuncttrue
\mciteSetBstMidEndSepPunct{\mcitedefaultmidpunct}
{\mcitedefaultendpunct}{\mcitedefaultseppunct}\relax
\EndOfBibitem
\bibitem[Hohm and Thakkar(2012)Hohm, and Thakkar]{Hohm2012}
Hohm,~U.; Thakkar,~A.~J. New {{Relationships Connecting}} the {{Dipole
  Polarizability}}, {{Radius}}, and {{Second Ionization Potential}} for
  {{Atoms}}. \emph{J. Phys. Chem. A} \textbf{2012}, \emph{116}, 697--703\relax
\mciteBstWouldAddEndPuncttrue
\mciteSetBstMidEndSepPunct{\mcitedefaultmidpunct}
{\mcitedefaultendpunct}{\mcitedefaultseppunct}\relax
\EndOfBibitem
\bibitem[Sarkisov \latin{et~al.}(2006)Sarkisov, Beigman, Shevelko, and
  Struve]{Sarkisov2006}
Sarkisov,~G.~S.; Beigman,~I.~L.; Shevelko,~V.~P.; Struve,~K.~W. Interferometric
  Measurements of Dynamic Polarizabilities for Metal Atoms Using Electrically
  Exploding Wires in Vacuum. \emph{Phys. Rev. A} \textbf{2006}, \emph{73},
  042501\relax
\mciteBstWouldAddEndPuncttrue
\mciteSetBstMidEndSepPunct{\mcitedefaultmidpunct}
{\mcitedefaultendpunct}{\mcitedefaultseppunct}\relax
\EndOfBibitem
\bibitem[Ma \latin{et~al.}(2015)Ma, Indergaard, Zhang, Larkin, Moro, and {de
  Heer}]{Ma2015}
Ma,~L.; Indergaard,~J.; Zhang,~B.; Larkin,~I.; Moro,~R.; {de Heer},~W.~A.
  Measured Atomic Ground-State Polarizabilities of 35 Metallic Elements.
  \emph{Phys. Rev. A} \textbf{2015}, \emph{91}, 010501\relax
\mciteBstWouldAddEndPuncttrue
\mciteSetBstMidEndSepPunct{\mcitedefaultmidpunct}
{\mcitedefaultendpunct}{\mcitedefaultseppunct}\relax
\EndOfBibitem
\bibitem[Dyugaev and Lebedeva(2016)Dyugaev, and Lebedeva]{Dyugaev2016}
Dyugaev,~A.~M.; Lebedeva,~E.~V. New Qualitative Results of the Atomic Theory.
  \emph{Jetp Lett.} \textbf{2016}, \emph{104}, 639--644\relax
\mciteBstWouldAddEndPuncttrue
\mciteSetBstMidEndSepPunct{\mcitedefaultmidpunct}
{\mcitedefaultendpunct}{\mcitedefaultseppunct}\relax
\EndOfBibitem
\bibitem[Ernst \latin{et~al.}(2016)Ernst, Santos, and Macchi]{Ernst2016}
Ernst,~M.; Santos,~L. H. R.~D.; Macchi,~P. Optical Properties of Metal--Organic
  Networks from Distributed Atomic Polarizabilities. \emph{CrystEngComm}
  \textbf{2016}, \emph{18}, 7339--7346\relax
\mciteBstWouldAddEndPuncttrue
\mciteSetBstMidEndSepPunct{\mcitedefaultmidpunct}
{\mcitedefaultendpunct}{\mcitedefaultseppunct}\relax
\EndOfBibitem
\bibitem[Gould(2016)]{Gould2016b}
Gould,~T. How Polarizabilities and {{C6}} Coefficients Actually Vary with
  Atomic Volume. \emph{J. Chem. Phys.} \textbf{2016}, \emph{145}, 084308\relax
\mciteBstWouldAddEndPuncttrue
\mciteSetBstMidEndSepPunct{\mcitedefaultmidpunct}
{\mcitedefaultendpunct}{\mcitedefaultseppunct}\relax
\EndOfBibitem
\bibitem[Bezchastnov \latin{et~al.}(2010)Bezchastnov, Pernpointner, Schmelcher,
  and Cederbaum]{Bezchastnov2010}
Bezchastnov,~V.~G.; Pernpointner,~M.; Schmelcher,~P.; Cederbaum,~L.~S.
  Nonadditivity and anisotropy of the polarizability of clusters:
  {Relativistic} finite-field calculations for the {Xe} dimer. \emph{Phys. Rev.
  A} \textbf{2010}, \emph{81}, 062507\relax
\mciteBstWouldAddEndPuncttrue
\mciteSetBstMidEndSepPunct{\mcitedefaultmidpunct}
{\mcitedefaultendpunct}{\mcitedefaultseppunct}\relax
\EndOfBibitem
\bibitem[A.~Manz \latin{et~al.}(2019)A.~Manz, Chen, J.~Cole, Gabaldon~Limas,
  and Fiszbein]{A.Manz2019}
A.~Manz,~T.; Chen,~T.; J.~Cole,~D.; Gabaldon~Limas,~N.; Fiszbein,~B. New
  Scaling Relations to Compute Atom-in-Material Polarizabilities and Dispersion
  Coefficients: Part 1. {{Theory}} and Accuracy. \emph{RSC Adv.} \textbf{2019},
  \emph{9}, 19297--19324\relax
\mciteBstWouldAddEndPuncttrue
\mciteSetBstMidEndSepPunct{\mcitedefaultmidpunct}
{\mcitedefaultendpunct}{\mcitedefaultseppunct}\relax
\EndOfBibitem
\bibitem[Bromley and Mitroy(2002)Bromley, and Mitroy]{Bromley2002b}
Bromley,~M. W.~J.; Mitroy,~J. Positron and Positronium Interactions with
  {{Cu}}. \emph{Phys. Rev. A} \textbf{2002}, \emph{66}, 062504\relax
\mciteBstWouldAddEndPuncttrue
\mciteSetBstMidEndSepPunct{\mcitedefaultmidpunct}
{\mcitedefaultendpunct}{\mcitedefaultseppunct}\relax
\EndOfBibitem
\bibitem[Henderson \latin{et~al.}(1997)Henderson, Curtis, Matulioniene, Ellis,
  and Theodosiou]{Henderson1997}
Henderson,~M.; Curtis,~L.~J.; Matulioniene,~R.; Ellis,~D.~G.; Theodosiou,~C.~E.
  Lifetime Measurements in {{Tl III}} and the Determination of the Ground-State
  Dipole Polarizabilities for {{Au I--Bi V}}. \emph{Phys. Rev. A}
  \textbf{1997}, \emph{56}, 1872--1878\relax
\mciteBstWouldAddEndPuncttrue
\mciteSetBstMidEndSepPunct{\mcitedefaultmidpunct}
{\mcitedefaultendpunct}{\mcitedefaultseppunct}\relax
\EndOfBibitem
\bibitem[Schwerdtfeger \latin{et~al.}(2000)Schwerdtfeger, Brown, Laerdahl, and
  Stoll]{Schwerdtfeger2000}
Schwerdtfeger,~P.; Brown,~J.~R.; Laerdahl,~J.~K.; Stoll,~H. The Accuracy of the
  Pseudopotential Approximation. {{III}}. {{A}} Comparison between
  Pseudopotential and All-Electron Methods for {{Au}} and {{AuH}}. \emph{J.
  Chem. Phys.} \textbf{2000}, \emph{113}, 7110--7118\relax
\mciteBstWouldAddEndPuncttrue
\mciteSetBstMidEndSepPunct{\mcitedefaultmidpunct}
{\mcitedefaultendpunct}{\mcitedefaultseppunct}\relax
\EndOfBibitem
\bibitem[Sarkisov(2022)]{Sarkisov2022}
Sarkisov,~G.~S. Laser Measurements of Static and Dynamic Dipole Polarizability
  for 11 Metal Atoms Using Fast Exploding Wires in Vacuum and Integrated-Phase
  Technique. \emph{Phys. Plasmas} \textbf{2022}, \emph{29}, 073502\relax
\mciteBstWouldAddEndPuncttrue
\mciteSetBstMidEndSepPunct{\mcitedefaultmidpunct}
{\mcitedefaultendpunct}{\mcitedefaultseppunct}\relax
\EndOfBibitem
\bibitem[Pašteka \latin{et~al.}(2017)Pašteka, Eliav, Borschevsky, Kaldor, and
  Schwerdtfeger]{Pašteka2017}
Pašteka,~L.; Eliav,~E.; Borschevsky,~A.; Kaldor,~U.; Schwerdtfeger,~P.
  Relativistic {Coupled} {Cluster} {Calculations} with {Variational} {Quantum}
  {Electrodynamics} {Resolve} the {Discrepancy} between {Experiment} and
  {Theory} {Concerning} the {Electron} {Affinity} and {Ionization} {Potential}
  of {Gold}. \emph{Physical Review Letters} \textbf{2017}, \emph{118}, 023002,
  Publisher: American Physical Society\relax
\mciteBstWouldAddEndPuncttrue
\mciteSetBstMidEndSepPunct{\mcitedefaultmidpunct}
{\mcitedefaultendpunct}{\mcitedefaultseppunct}\relax
\EndOfBibitem
\bibitem[Thierfelder and Schwerdtfeger(2010)Thierfelder, and
  Schwerdtfeger]{Thierfelder2010}
Thierfelder,~C.; Schwerdtfeger,~P. Quantum electrodynamic corrections for the
  valence shell in heavy many-electron atoms. \emph{Physical Review A}
  \textbf{2010}, \emph{82}, 062503, Publisher: American Physical Society\relax
\mciteBstWouldAddEndPuncttrue
\mciteSetBstMidEndSepPunct{\mcitedefaultmidpunct}
{\mcitedefaultendpunct}{\mcitedefaultseppunct}\relax
\EndOfBibitem
\end{mcitethebibliography}
\end{document}